\documentclass[aps,prd,twocolumn,showpacs,superscriptaddress,preprintnumbers, floatfix,nofootinbib]{revtex4-1}
\usepackage{amsmath}
\usepackage{hyperref, url}
\usepackage{graphicx,subfigure}
\usepackage[normalem]{ulem}
\usepackage{epsfig}
\usepackage{color,soul}
\usepackage{multirow}
\usepackage{array}
\usepackage[utf8]{inputenc}
\usepackage[T1]{fontenc}
\usepackage[dvipsnames]{xcolor}
\usepackage{comment}
\usepackage{aas_macros}
\hypersetup{colorlinks   = true,
            urlcolor     = blue,
            citecolor    = blue,
            linkcolor    = blue,
            menucolor    = blue,
            anchorcolor  = blue,
            filecolor    = blue}

\enlargethispage{\baselineskip}

\usepackage{graphicx}

\newcommand{\Mpl}{M_{\text{pl}}}

\newcommand{\nn}{\nonumber}

 \newcommand{\Ocal}{{\mathcal O}}
 
 \newcommand{\Lcal}{{\mathcal L}}

\begin{document}
\preprint{IPMU23-0024, KEK-QUP-2023-0014, KEK-TH-2534, KEK-Cosmo-0316}

\title{Detection of Bosenovae with Quantum Sensors on Earth and in Space}

\author{Jason Arakawa}
\email{arakawaj@udel.edu}
\affiliation{Department of Physics and Astronomy, University of Delaware, Newark, Delaware 19716, USA}

\author{Joshua Eby}
\email{joshaeby@gmail.com}
\affiliation{Kavli Institute for the Physics and Mathematics of the Universe (WPI), \mbox{The University of Tokyo Institutes for Advanced Study, The University of Tokyo, Kashiwa, Chiba 277-8583, Japan}}

\author{Marianna S. Safronova}
\email{msafrono@udel.edu}
\affiliation{Department of Physics and Astronomy, University of Delaware, Newark, Delaware 19716, USA}
\affiliation{Joint Quantum Institute, National Institute of Standards and Technology and the University of Maryland, College Park, Maryland 20742, USA}

\author{Volodymyr Takhistov}
\email{vtakhist@post.kek.jp}
\affiliation{International Center for Quantum-field Measurement Systems for Studies of the Universe and Particles (QUP, WPI),
High Energy Accelerator Research Organization (KEK), Oho 1-1, Tsukuba, Ibaraki 305-0801, Japan}
\affiliation{Theory Center, Institute of Particle and Nuclear Studies (IPNS), High Energy Accelerator Research Organization (KEK), Tsukuba 305-0801, Japan
}
\affiliation{Kavli Institute for the Physics and Mathematics of the Universe (WPI), \mbox{The University of Tokyo Institutes for Advanced Study, The University of Tokyo, Kashiwa, Chiba 277-8583, Japan}}

\author{Muhammad H. Zaheer}
\email{hani@udel.edu}
\affiliation{Department of Physics and Astronomy, University of Delaware, Newark, Delaware 19716, USA}

\date{\today}

\begin{abstract}
In a broad class of theories, the accumulation of ultralight dark matter (ULDM) with particles of mass $10^{-22}~\textrm{eV} < m_{\phi} < 1~\textrm{eV}$ leads the to formation of long-lived bound states known as boson stars. When the ULDM exhibits self-interactions, prodigious bursts of energy carried by relativistic bosons are released from collapsing boson stars in bosenova explosions. We extensively explore the potential reach of terrestrial and space-based experiments for detecting transient signatures of emitted relativistic bursts of scalar particles, including ULDM coupled to photons, electrons, and gluons, capturing a wide range of motivated theories. For the scenario of relaxion ULDM, we demonstrate that upcoming experiments and technology such as nuclear clocks as well as space-based interferometers will be able to sensitively probe orders of magnitude in the ULDM coupling-mass parameter space, challenging to study otherwise, by detecting signatures of transient bosenova events. Our analysis can be readily extended to different scenarios of relativistic scalar particle emission.
\end{abstract}

\maketitle

\section{Introduction}
\label{sec:introduction}

The influence of dark matter (DM), which constitutes $\sim85\%$ of matter in the Universe, has been firmly established by astronomical observations (see e.g. \cite{Bertone:2004pz} for a review). With decades of searches for its non-gravitational interactions, the nature of DM remains mysterious. A particularly well-studied scenario has been that of Weakly Interacting Massive Particles (WIMPs), with typical masses in the range of 1 GeV -- 100 TeV, often associated with theories attempting to address the hierarchy problem. A well-motivated DM paradigm that has been less explored is ultralight dark matter (ULDM) \cite{ULDM}, where extremely low-mass (\(10^{-22}~\text{eV}\lesssim m_{\phi} \lesssim 1~ \rm{eV}\)) bosonic fields $\phi$ behave as classical waves due to their high occupation numbers and large de-Broglie wavelengths.

There are a multitude of ways in which ULDM can interact with the SM, which has lead to recent development of a broad and diverse experimental search program~\cite{ULDM,Antypas:2022asj,adams2023axion}. 
Due to the wide range of possible masses $m_\phi$ of ULDM, a large number of experiments, including atomic clocks \cite{Beloy:2020tgz, Filzinger:2023zrs, Sherrill:2023zah, 2022PhRvL.129x1301K, Hees:2016gop, Aharony:2019iad}, optical cavities \cite{Kennedy:2020bac, Tretiak:2022ndx, Campbell:2020fvq}, optical interferometers \cite{Savalle:2020vgz, Aiello:2021wlp}, spectroscopy \cite{Oswald:2021vtc, Zhang:2022ewz, VanTilburg:2015oza}, tests of the equivalence principle (EP) \cite{Hees:2018fpg,Berge:2017ovy}, mechanical resonators \cite{Branca:2016rez}, fifth force searches \cite{Fischbach:1996eq, Murata:2014nra}, gravitational wave (GW) detectors \cite{Vermeulen:2021epa, Fukusumi:2023kqd}, and torsion balance experiments \cite{Adelberger:2003zx} are currently exploring the ULDM parameter space. Scalar DM that couples to the SM can cause tiny variations in the fundamental constants due to the coherent oscillations of the DM background, providing key observables that have been extensively studied \cite{Arvanitaki:2014faa, Antypas:2022asj}.

ULDM can generally coalesce into long-lived bound structures known as \emph{boson stars} \cite{Kaup:1968zz,Ruffini:1969qy,Colpi:1986ye}. Such states can form through purely gravitational interactions \cite{Levkov:2018kau,Chen:2020cef} or through self-interactions \cite{Kirkpatrick:2020fwd,Chen:2021oot,Kirkpatrick:2021wwz}, and at small-enough densities they are stable both under perturbations \cite{Seidel:1990jh,Chavanis:2011zi,Chavanis:2011zm,Eby:2014fya} and decay \cite{Hertzberg:2010yz,Eby:2015hyx,Mukaida:2016hwd,Braaten:2016dlp,Eby:2017azn,Zhang:2020bec,Hertzberg:2020xdn,Eby:2020ply}. The full gamut of phenomenological implications and experimental signatures of boson stars has been poorly explored.

A stable boson star configuration can be understood as a balance between the (attractive) self-gravity of the ULDM particles and their (repulsive) gradient energy. This balance can be sustained as long as the star is relatively dilute, such that the contribution of self-interactions is negligible. However, if the mass of the star grows through merger events \cite{Mundim:2010hi,Cotner:2016aaq,Schwabe:2016rze,Eby:2017xaw,Hertzberg:2020dbk,Du:2023jxh} and/or accretion of ULDM from the background \cite{Chen:2020cef,Chan:2022bkz,Dmitriev:2023ipv}, eventually the self-interactions contribute and, if they are attractive in nature, they can destabilize the star.

When the boson star begins to collapse, its density rapidly increases, as does the binding energy of the collapsing ULDM particles. As the size of the star approaches the Compton wavelength of the ULDM, $2\pi/m_\phi$, annihilations of the ULDM particles to relativistic ones rapidly deplete the energy of the collapsing star. This typically occurs far before the star reaches its Schwarzschild radius and therefore the boson star does not collapse\footnote{This discussion holds when the boson star is still non-relativistic when the self-interactions become relevant. Note on the contrary that, if the self-interactions are exceptionally weak, then the boson star can become a black hole directly as a result of mass increase. See discussion in Section~\ref{sec:bosonstar}. 
In the special case of ULDM axions with Planck-scale decay constants, these conclusions can also be modified~\cite{Helfer:2016ljl,Eby:2020ply}.} to a black hole \cite{Eby:2016cnq,Levkov:2016rkk,Eby:2017xrr}.
The pressure resulting from the relativistic emission reverses the collapse and, after a few collapse/emission cycles \cite{Levkov:2016rkk}, a large fraction of the boson star energy is emitted and the cold ULDM that remains settles into a dilute, gravitationally-bound configuration again. This explosion of relativistic ULDM that occurs at the endpoint of boson star collapse is known as \emph{bosenova}.

The bosenovae are a promising target for distinct direct ULDM searches.
In previous work \cite{Eby:2021ece}, some of us studied the bosenova signal in detail for the case of ULDM composed of axion-like (parity-odd) fields, finding that current and near-future experiments, e.g. those using LC-circuit-based detectors \cite{Sikivie:2013laa,Kahn:2016aff} like ABRACADABRA \cite{Ouellet:2018beu,Salemi:2021gck}, SHAFT \cite{Gramolin:2020ict} and DM-Radio \cite{Silva-Feaver:2016qhh,Rapidis:2022gti,DMRadio:2023igr}, could viably search for relativistic ULDM bursts from collapsing boson stars. 
This was possible both because the energy density in the burst could exceed the background density of DM, $\rho_{\rm DM} \simeq 0.4$ GeV/cm$^3$, and because the wave spreading of the burst in flight led to a signal that was highly coherent over the duration of the burst in the detector.

As stressed in Ref.~\cite{Eby:2021ece}, axion-like fields generically exhibit a close relationship between the self-interaction coupling $\lambda$ of the ULDM field and its couplings to the SM: both couplings are typically determined by the same scale $f_a$, the axion decay constant. While such relationship is predictive, other scalar fields (e.g. parity-even fields or dilatons) can have independent distinct DM-SM and self-interaction couplings. This highlights the importance of general exploration of possible scenarios for scalar field bursts.

In this paper, building on previous work \cite{Eby:2021ece}, we consider and develop a comprehensive methodology to analyze a general ultralight scalar (parity-even) field, and determine the prospects for detecting transient signals of relativistic scalars emitted in bosenovae. We describe the DM-SM interactions in effective field theory (EFT) with dilatonic couplings, e.g. with couplings to photons, electrons, and gluons, capturing a large swath of possible theories. Although most of the analysis is model-independent, we also examine in detail the specific case of a \emph{relaxion} field \cite{Graham:2015cka} as a concrete realization of the scalar DM with dilatonic couplings \cite{Banerjee:2018xmn,Chatrchyan:2022dpy}. 

The paper is organized as follows. Section \ref{sec:EFT} outlines the leading-order couplings between the DM and the SM within the framework of EFT. We also include a detailed discussion of constraints on the self-interactions of ULDM. We proceed by describing boson stars and the properties of their explosions in Section \ref{sec:bosonstar}. Section \ref{sec:detection} discusses the detection prospects of the transient scalar bursts, and Section \ref{sec:results} presents the results, including general sensitivity analyses for photon, electron, and gluon couplings. We also analyze detection prospects in the scenario of relaxion DM. We reserve Section \ref{sec:conclusion} for our conclusions and future outlook. We work in natural units throughout with $\hbar=c=1$.

\section{Effective Couplings}
\label{sec:EFT}

\subsection{Dilatonic ULDM-SM couplings}
 
We consider the case of DM as composed of a real scalar field \(\phi\). 
Direct scalar couplings of $\phi$ with the SM 
can lead to effective variation of fundamental constants, which are targets for direct searches (e.g. in quantum sensors) as discussed in the Introduction (see e.g.~\cite{Hees:2018fpg}).
Such operators can be characterized within EFT as
\begin{align}
    \mathcal{L}_{\text{int}} = \sum_{n,i} d^{(n)}_i \bigg(\frac{\phi}{\Mpl}\bigg)^n \mathcal{O}^i_{\text{SM}}\,,
\end{align}
where \(d^{(n)}_i\) are the dilatonic couplings labeled by an index \(i\) and $\phi$-exponent \(n\), and \(\Ocal^i_{\rm{SM}}\) are corresponding operators within the SM.
The leading 
effective couplings, which 
appear at linear order ($n=1$), are
\begin{align} \label{eq:Lint}
    \mathcal{L}_{\text{int}} \supset \frac{\sqrt{4\pi}\,\phi}{\Mpl}
    \bigg( &d_{m_e}^{(1)}m_e \overline{e} e 
        + \frac{d_{e}^{(1)}}{4} F_{\mu \nu}F^{\mu \nu} \nn \\
        &+ \frac{d_g^{(1)} \beta(g_s)}{2g_s}G_{\mu\nu}^a G^{a\mu\nu}\bigg)\,,
\end{align}
where $e$ is the electron field with mass $m_e$, $F^{\mu\nu}$ and $G^{a\mu\nu}$ are the photon and gluon field strength tensors (respectively), and $g_s$ and $\beta(g_s)$ are the coupling and beta function of QCD.
The three interaction terms above lead to effective variation of fundamental quantities (respectively): the electron-to-proton mass ratio \(m_e/m_p\), the fine-structure constant  \(\alpha\), and the strong coupling constant $\alpha_s$.

In this work, we focus on the linear-in-$\phi$ couplings given in Eq.~\eqref{eq:Lint}. However, we note that a family of operators of higher dimensionality exists as well. The next-to-leading-order couplings are quadratic in $\phi$, and can have interesting phenomenological consequences~\cite{Hees:2018fpg,Banerjee:2022sqg,Bouley:2022eer}. A qualitative difference between linear and quadratic couplings is that the latter can lead to screening (or anti-screening)
of the field in the presence of significant matter densities (see \cite{Hees:2018fpg} for details). This could suppress (or enhance) detection sensitivities for terrestrial experiments, even when the detection sensitivity is otherwise dominated by the leading-order coupling. This additional model dependence provides one motivation for the use of space-based experiments \cite{Banerjee:2022sqg}.

\subsection{ULDM self-interactions}

In addition to the ULDM-SM interaction Lagrangian of Eq.~\eqref{eq:Lint}, we parameterize the $\phi$-only Lagrangian 
as
\begin{align} \label{eq:Lphi}
    \mathcal{L}_\phi &= \frac{1}{2}\partial_{\mu}\phi \partial^{\mu} \phi - \frac{1}{2}m_{\phi}^2 \phi^2 
    + \frac{\lambda}{4!}\phi^4 + ...
\end{align}
Generically, $\lambda$ 
can be positive or negative, corresponding to attractive and repulsive self-interactions (respectively). We focus on the case of attractive self-interactions, as the collapse of boson stars (which gives rise to the signal we are investigating) only arises when $\lambda > 0$ \cite{Chavanis:2011zi,Chavanis:2011zm,Eby:2014fya}. Attractive self-interactions can generically be realized within ultraviolet (UV) completions (see e.g. \cite{Fan:2016rda}).

We note that a cubic term, $-\lambda_3\phi^3/3!$,
is possible as well, and can be attractive or repulsive. However, its leading contribution is to mediate $2\to 1$ or $1\to2$ transitions which are suppressed in the non-relativistic limit relevant to DM studies. For simplicity, in this work we set $\lambda_3=0$ and focus on $\lambda$. In boson star collapse, the presence of this cubic term or terms proportional to $\phi^{n}$ with $n > 4$ may modify the emission spectrum of bosons in the bosenova, which is a topic we leave for future work. See Section \ref{sec:bosonstar} for a brief discussion.

The couplings considered in this work are extremely small, $\lambda\ll 1$, as motivated by well-studied models of ULDM. As a benchmark example, axion-like fields often have self-interaction couplings determined by their mass $m_\phi$ and symmetry-breaking scale $f_a$ as $\lambda \simeq (m_\phi/f_a)^2$. For QCD axions, one has $m_\phi f_a\simeq \Lambda_{\rm QCD}^2\simeq (75$ MeV$)^2$ (the QCD scale, see e.g. \cite{GrillidiCortona:2015jxo}), implying $\lambda \simeq m_\phi^4/\Lambda_{\rm QCD}^4 \simeq 3\times10^{-56}(m_\phi/\mu {\rm eV})^4$. Studies of so-called fuzzy DM \cite{Press:1989id,Sin:1992bg,Hu:2000ke,Peebles:2000yy,Amendola:2005ad} often use a benchmark of $m_\phi\simeq 10^{-22}$ eV and $f_a \simeq 10^{16}$ GeV, which gives $\lambda \simeq 10^{-94}$ (see \cite{Hui:2016ltb,Ferreira:2020fam} for recent reviews). We will discuss constraints on $\lambda$ in the next section.

Note that for axion-like fields, the SM couplings are typically connected to the same high scale $f_a$. For example, the linear axion-photon interaction is typically parameterized as $\Lcal_{a\gamma} \propto (\phi/f_a)F^{\mu\nu}\tilde{F}_{\mu\nu}$, where $\tilde{F}$ is the dual field strength of the photon. This relation is not applicable to the scalars considered in this work.

\subsection{Constraints on $\lambda$}
\label{ssec:lambdaconstraints}

In parameterizing the scalar field EFT, we treat \(\lambda\) as independent of the SM couplings at the Lagrangian level (in contrast to the axion case just described).
Therefore, it is useful to explore general constraints 
on \(\lambda\), which we describe below and summarize in Fig.~\ref{fig:lambda_constraints}.

\subsubsection{Bullet Cluster}

\begin{figure}
    \centering
    \includegraphics[width = \linewidth]{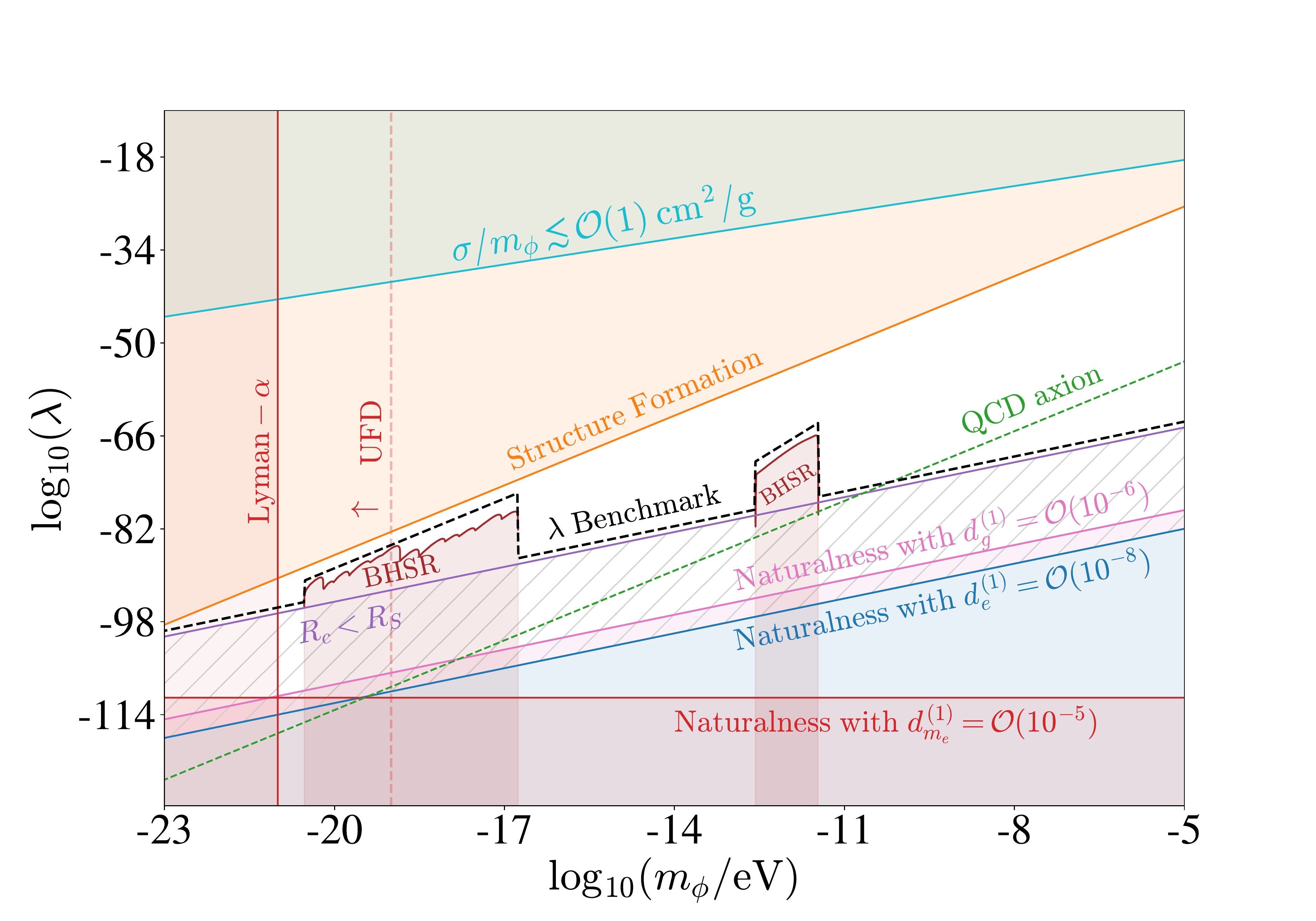}
    \caption{Summary of the parameter space relevant for the bosenova signal analyzed in this work. Constraints on \(\lambda\) illustrated arise from the Bullet Cluster 
    (cyan) \cite{Markevitch:2003at}, structure formation (orange) \cite{Cembranos:2018ulm}, Lyman-$\alpha$ \cite{Irsic:2017yje,Rogers:2020ltq} and ultra-faint dwarf (UFD) galaxies \cite{Dalal:2022rmp} (red vertical thick and dashed lines, respectively), black hole superradiance~\cite{Baryakhtar:2020gao,Unal:2020jiy} (BHSR, brown), as well as the lower bounds for naturalness arising from \(d_{e}\) (blue), \(d_{m_e}\) (red), and \(d_g\) (pink). The naturalness 
    lines assume benchmark values for the couplings which are allowed by current experiments, as labeled. We also illustrate the region where critical boson stars would 
    not undergo bosenova due to general relativistic effects (purple hatched region).
    The dashed black line represents the benchmark values for $\lambda$ that we choose at each mass, as given in Eq.~\eqref{eq:lambdaB} and explained in the Section~\ref{sec:results}. Finally, for comparison, we show the QCD axion line corresponding to $|\lambda|\simeq m_\phi^4/\Lambda_{\rm QCD}^4$ (green dotted line); see text for details.}
    \label{fig:lambda_constraints}
\end{figure}

The coupling $\lambda$ can be constrained by 
the observed gross distribution of DM in galaxy-cluster collisions, including the Bullet Cluster \cite{Markevitch:2003at}. The tree-level scattering cross section $\sigma$ for \(\phi-\phi\) scattering via $\lambda$ is given by
\begin{align}
    \frac{\sigma}{m_{\phi}} = \frac{\lambda^2}{16\pi m_{\phi}^3}\,.
\end{align}
Gravitational-lensing observations of the Bullet Cluster constrain the DM self-interaction cross section to be \cite{Markevitch:2003at}
\begin{equation} \label{eq:bulletcluster}
    \frac{\sigma}{m_\phi} \lesssim \mathcal{O}(1) ~\text{cm}^2/\text{g}\,,
\end{equation}
which implies
\begin{align}
    \lambda \lesssim  10^{-11} \bigg(\frac{m_{\phi}}{\rm eV}\bigg)^{3/2}\,.
\end{align}
This model-independent constraint on DM self-interactions is illustrated by the cyan shaded region in Fig.~\ref{fig:lambda_constraints}.

\subsubsection{Cosmological constraints}
\label{ssec:LSS}

The presence of ultralight fields of mass $m$ and self-interaction coupling $\lambda$ gives rise to contributions to the evolution of the matter power spectrum, as discussed in e.g. \cite{Arvanitaki:2014faa,Fan:2016rda}. Density perturbations $\delta_{\bf k} \equiv (\rho_{\bf k}-\bar{\rho})/\bar{\rho}$ at scales $|{\bf k}|$ will evolve as
\begin{equation} \label{eq:deltadot}
    \ddot\delta_{\bf k} + 2H\dot\delta_{\bf k} \simeq \left[\frac{4\pi\bar{\rho}}{\Mpl^2} - \left(\frac{{\bf k}^2}{4m_\phi^2 a^2} - \frac{3\lambda\phi^2}{16m_\phi^2}\right)\frac{{\bf k}^2}{a^2}\right]\delta_{\bf k},
\end{equation}
where $H$ is the Hubble parameter, $a$ is the scale factor of the universe, and $\bar\rho$ and $\rho_{\bf k}$ are the the average DM density and the density at scale $|{\bf k}|$ (respectively). 

The first term in parentheses on the right-hand side of Eq.~\eqref{eq:deltadot} contributes to a suppression of the matter power spectrum on a distance scale of $L_m \simeq [\pi^3\Mpl^2/(\bar{\rho} m_\phi^2)]^{1/4}$.
In the mass range $m_\phi \lesssim 10^{-21}-10^{-20}$ eV, this can lead to constraints from measurements of small-scale structure in Lyman-$\alpha$ forest \cite{Irsic:2017yje,Rogers:2020ltq}.
A stronger constraint, approximately $m_\phi \gtrsim 10^{-19}$ eV, was recently derived in Ref.~\cite{Dalal:2022rmp} by modeling stellar velocity dispersion in ultra-faint dwarf (UFD) galaxies. The result depends on astrophysical assumptions about the evolution and tidal stripping of DM in particular UFDs (see e.g. \cite{DuttaChowdhury:2023qxg}). 
These constraints are illustrated by the red thick and dashed lines (respectively) in Fig.~\ref{fig:lambda_constraints}.

The second term in parentheses on the right-hand side of Eq.~\eqref{eq:deltadot} suppresses (enhances) the matter power spectrum when $\lambda < 0$ ($\lambda>0$), corresponding to a repulsive (attractive) interaction; the relevant scale is approximately $L_\lambda \simeq [3\pi|\lambda|\Mpl^2/(8m_\phi^4)]^{1/2}$. Observational constraints require 
$L_\lambda \lesssim$  Mpc \cite{Cembranos:2018ulm}, corresponding to
\begin{equation}
    \lambda \lesssim 3\times 10^{-79} \bigg(\frac{m_{\phi}}{10^{-18}~\text{eV}}\bigg)^4\,.
\end{equation}
We show this 
constraint in orange color in Fig.~\ref{fig:lambda_constraints}. 

Note finally that the constraints of this section and the previous one assume that $\phi$ constitutes an $\Ocal(1)$ fraction of the DM of the universe, and vanish when this is not so.

\subsubsection{Naturalness of $\lambda$}

\begin{figure}
    \centering
    \includegraphics[width = .65\linewidth]{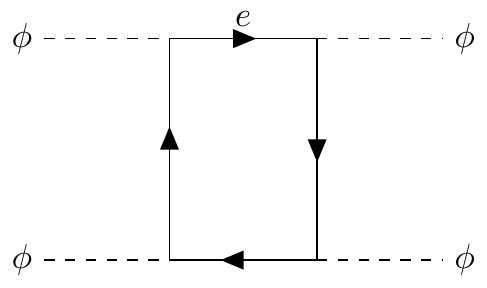}
    \hspace{.4cm}
    \includegraphics[width = .65\linewidth]{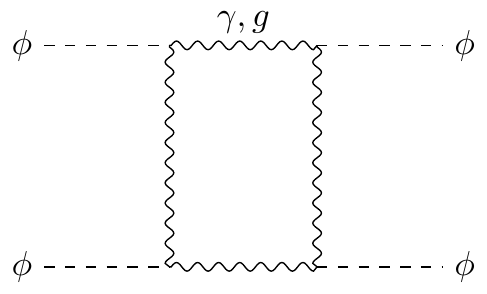}
    \caption{One-loop diagrams that generate corrections to \(\lambda\) from the linear \(\phi\) coupling \(d_{m_e}^{(1)}\) (upper panel) or \(d_{e}^{(1)}\) and \(d_{g}^{(1)}\) (lower). }
    \label{fig:linloop}
\end{figure}

We also consider theoretical constraints on \(\lambda\), demanding that the values of \(\lambda\) considered do not require technical fine-tuning of the theory (see e.g.~\cite{Craig:2022uua} for a recent summary). 
The $\phi$-SM couplings can generate self-interactions proportional to $\phi^4$ through box diagrams, illustrated in Fig.~\ref{fig:linloop} for electron (upper) or gauge boson (lower) couplings. 
If this contribution is much larger than the physical, observable $\lambda$, then this implies an ``unnatural'' cancellation between the loop contribution and the bare Lagrangian parameter, which are \emph{a priori} unrelated. 
This allows us to set 
an approximate lower bound on the value on \(\lambda\), 
essentially by the requirement that the effective $\lambda$ coupling is not much smaller than the one generated at one-loop. 

Integrating the box diagrams in Fig.~\ref{fig:linloop} up to some UV cutoff $\Lambda_{\rm UV}$ gives
\begin{align} \label{eq:lambdaUV}
    \lambda(\Lambda_{\text{UV}}) &\simeq \lambda(\mu_{0}) - \frac{(d^{(1)}_{m_e})^4 m_e^4}{8\pi^2 \Mpl^4} \log{\bigg(\frac{\Lambda_{\text{UV}}}{\mu_{0}}\bigg)}\nonumber\\
    &+ \frac{(d^{(1)}_{e})^4}{32\pi^2 \Mpl^4} \Lambda_{\text{UV}}^4 + \frac{(d^{(1)}_{g} \beta(g_s))^4}{2\pi^2 g_s^4 \Mpl^4} \Lambda_{\text{UV}}^4\,,
\end{align}
where $\mu_0$ is some reference energy scale. Eq.~\eqref{eq:lambdaUV} represents a rough estimate for illustration only; the gluon coupling in particular neglects any color factors that may alter the expression by an \(\Ocal(1)\) factor.

In its current form, each term of Eq.~(\ref{eq:lambdaUV}) has two unknown parameters: $d_{i}^{(1)}$ and $\Lambda_{\text{UV}}$. To obtain a definitive bound on $\lambda$ from this expression, we also need a constraint on $\Lambda_{\text{UV}}$, especially for the gauge boson loops, which are quartic in $\Lambda_{\text{UV}}$. As noted in Ref.~\cite{Bouley:2022eer}, there is an upper bound on the UV cutoff of the theory from requiring that the one-loop mass corrections induced by the DM-SM interactions are also natural. Since the interactions involved in the corrections to $m_{\phi}^2$ and $\lambda$ are the same, the UV scales that arise in loop integration are the same. The loops have quadratic and quartic UV divergences, which means that the UV cutoff must be sufficiently small for the mass corrections to be subdominant to the tree-level mass of \(\phi\) \cite{Banerjee:2022sqg}. By enforcing \(\delta(m_{\phi}^2) \lesssim m_{\phi}^2\), we obtain
\begin{align}
    \Lambda^{m_e}_{\text{UV}} 
        &\lesssim 
        2^{3/2}\frac{\pi m_{\phi}\Mpl}{d^{(1)}_{m_e}m_e}\,, \\
    \Lambda_{\text{UV}}^{e} 
        &\lesssim 
        2^{5/4}\sqrt{\frac{\pi m_{\phi} \Mpl}{d^{(1)}_e}}\,,\\
    \Lambda_{\text{UV}}^{g} 
        &\lesssim 
        2^{3/4}\sqrt{\frac{\pi m_{\phi} \Mpl g_s }{d^{(1)}_g  \beta(g_s)}}\,,
\end{align}
\noindent for the electron, photon, and gluon couplings respectively. 
Substituting these limits on the UV cutoff into the renormalized \(\lambda\) in Eq.~\eqref{eq:lambdaUV}, we find 
\begin{align}
    \lambda &\gtrsim \frac{(d_{m_e}^{(1)})^4 m_{e}^4}{8\pi^2 \Mpl^4} \log{\bigg(\frac{\sqrt{8} \pi m_{\phi}M_{{\text{pl}}}}{d_{m_e}\mu_{0}}\bigg)} \,, \\
    \lambda &\gtrsim (d_e^{(1)})^2 \frac{m^2_{\phi}}{M_{\text{pl}}^2} \,, \\
    \lambda &\gtrsim 2\,\bigg(d_g^{(1)}\frac{\beta(g_s)}{g_s}\bigg)^2\frac{m^2_{\phi}}{M_{\text{pl}}^2}\,,
\end{align}
for the electron, photon, and gluon couplings, respectively.

Therefore, the ``natural'' regions of $\lambda$ depend on the unknown SM couplings. For illustration, in Fig.~\ref{fig:lambda_constraints} we show the natural scale of $\lambda$ assuming $d_e^{(1)}=10^{-8}$,  $d_{m_e}^{(1)}=10^{-5}$, and $d_g = 10^{-6}$ (blue, red, and pink, respectively), which are benchmark values that are allowed by current constraints across the full range of $m_\phi$ we consider in this work (c.f. the results of Section \ref{sec:results}). For the electron coupling in Fig.~\ref{fig:lambda_constraints}, we assume \(\log(\sqrt{8} \pi m_{\phi}M_{{\text{pl}}}/d^{(1)}_{m_e}\mu_0) \sim \mathcal{O}(1)\), as the prefactor primarily determines its magnitude. For the gluon coupling, the $\beta$-function is given by $\beta(g_s)/g_s = -(11 - n_s/6 - 2n_f/3) \,g_s^2/16\pi^2$. Here we use $\alpha_s \equiv g_s^2/4\pi = 1$ to represent that QCD is non-perturbative at these energy scales. The results do not strongly depend on the value chosen for $\alpha_s$, and the 
$R_c > R_S$ constraint (purple region) discussed below, where $R_S$ is the Schwarzschild radius, is strictly stronger. Even if for a choice of larger benchmark couplings limited by the EP bounds, the $R_c > R_S$ constraint remains stronger.
Additionally, we do not specify a UV completion, which could modify $\beta(g_s)$ via the introduction of additional SU(3)$_c$ colored scalars ($n_s$) and fermions ($n_f$), but this also does not appreciably change the results shown here. For simplicity, we choose the SM contribution $n_s = 0$ and $n_f = 6$.

\subsubsection{Black hole superradiance}

In the vicinity of a rapidly rotating black hole,
ultralight scalars can be produced in hydrogen-like bound states by extracting energy and angular momentum from the black hole. This process, known as \emph{black hole superradiance}, is most efficient for scalars
whose Compton wavelength is comparable to the 
Schwarzschild radius of the black hole, i.e. $2\pi m_\phi^{-1} \simeq R_S \equiv 2 M_{\rm BH}/\Mpl^2$, where $M_{\rm BH}$ is the black hole mass~\cite{Arvanitaki:2010sy,Arvanitaki:2014wva,Arvanitaki:2016qwi}. 

Because superradiance causes the host to spin down over time, observations of black holes with large spins today have led to constraints on the existence of such scalars, in the mass range $m_\phi \simeq 10^{-13}-10^{-11}$~eV (for solar-mass black holes) \cite{Baryakhtar:2020gao} and $m_\phi \simeq 10^{-21} - 10^{-17}$~eV (for supermassive black holes) \cite{Unal:2020jiy}.
We illustrate these constraints by the brown lines in Fig.~\ref{fig:lambda_constraints}. These studies take advantage of a large number of measured black hole spins, which are difficult to measure and in some cases have $\Ocal(1)$ uncertainties, and are likely to improve in the future. Direct laboratory searches, including those studied here, are complementary to these indirect astrophysical limits. 

Self-interactions can also lead to important effects on black hole superradiance, as the constraints in Fig.~\ref{fig:lambda_constraints} illustrate. If the scalar field density around the black hole exceeds a critical value, attractive self-interactions destabilize the cloud, causing it to collapse and quenching its exponential growth~\cite{Arvanitaki:2010sy}. Depending on the value of the self-interaction coupling $\lambda$, the cloud can also transition to a steady-state configuration, where superradiant states are excited and relaxed at roughly equal rates~\cite{Baryakhtar:2020gao,Branco:2023frw}. This also quenches the growth of the cloud. The end result of both processes is that, for strong-enough self-interaction couplings, the superradiance rate is insufficient to spin down the black hole on astrophysical timescales, and the constraints are relaxed accordingly. Roughly, spin-down occurs within the lifetime of the universe when~\cite{Branco:2023frw}
\begin{equation}
    |\lambda| < 10^{-64}\,\alpha_g 
        \left(\frac{m_\phi}{10^{-10}\,{\rm eV}}\right)^{5/2}\,,
\end{equation}
though this condition is approximate and only valid when the gravitational coupling $\alpha_g \equiv M_{\rm BH} m_\phi/\Mpl^2$ is order unity for a given $M_{\rm BH}$. 
A full study of the spin-down rates gives rise to the upper edge of the superradiance constraints seen in~\cite{Baryakhtar:2020gao,Unal:2020jiy}, reproduced in Fig.~\ref{fig:lambda_constraints}.

\subsection{Relaxion dark matter}

As a concrete example of a well-motivated particle physics model that is captured by our analysis, consider the relaxion, a scalar field proposed to alleviate the hierarchy problem in the Higgs sector \cite{Graham:2015cka}. 
In these models, the cosmological rolling of the $\phi$ (relaxion) field scans the electroweak scale $v_{\rm ew}$ until a Higgs-dependent backreaction traps $\phi$ in a local minimum with a Higgs mass scale close to the measured value. Additionally, by displacing the relaxion field from the minimum of its potential, either by reheating the universe after inflation \cite{Banerjee:2018xmn} or through stochastic quantum fluctuations \cite{Chatrchyan:2022dpy}, the relaxion is able to acquire a significant energy density, making it a viable ULDM candidate.

Because the backreaction potential is generated through direct $\phi$-Higgs couplings, the relaxion naturally acquires Higgs-like interactions with SM fields parameterized by an effective scalar mixing angle $\theta_{h\phi}$ (see e.g. \cite{Flacke:2016szy} for details). We therefore characterize the effective relaxion couplings with a Lagrangian of the form
\begin{align}
    \mathcal{L} \supset \sin{\theta_{h\phi}}\frac{\phi}{\sqrt{2}}\bigg(\sum_f \frac{m_f}{v_{\text{ew}}}\overline{f}f &+ c_{\gamma} \frac{\alpha}{4\pi v_{\text{ew}}}F_{\mu\nu}F^{\mu\nu} \nn \\
    &+ c_g \frac{\alpha_s}{4\pi v_{\text{ew}}}G^a_{\mu\nu}G^{a\mu\nu} \bigg)\,.
\end{align}
where $f$ is a SM fermion field (we will take $f=e$ in what follows).
Comparing each term to the EFT parameterization in Eq.~\eqref{eq:Lint}, we can 
derive the relationship of $\sin\theta_{h\phi}$ to each of the dilatonic couplings $d_i^{(1)}$:
\begin{align}
\label{eq:relaxionmixing}
    \sin{\theta_{h\phi}} = \frac{v_{\rm ew}}{\Mpl}\times
    \begin{cases}
       \displaystyle \sqrt{8\pi} d_{m_e}^{(1)} & {\rm (electron)} \\
       \\
       \displaystyle \frac{\sqrt{8\pi^3}}{\alpha} d_e^{(1)} & {\rm (photon)} \\
       \\
       \displaystyle \frac{\sqrt{8\pi^2}\beta(g_s)}{\alpha_s^{3/2}} d_g^{(1)} & {\rm (gluon)}
    \end{cases}\,.
\end{align}

\noindent where for simplicity we took $c_\gamma=c_g=1$. As with axion-like fields, relaxion self-interactions can be parameterized by the effective coupling $\lambda\simeq m_\phi^2/f_a^2$. Nonetheless, taking $f_a$ is a free parameter, this implies that $\lambda$ can take a wide range of values in the allowed region of Fig.~\ref{fig:lambda_constraints}. 

\section{Boson Stars}
\label{sec:bosonstar}

ULDM can form self-gravitating bound states known as boson stars \cite{Kaup:1968zz,Ruffini:1969qy,Colpi:1986ye}, 
which can be understood as quasi-static standing-wave solutions of the low-energy equations of motion: the Gross-Pitaevskii--Poisson (GPP) equations \cite{Chavanis:2011zi,Chavanis:2011zm}. These equations are easily derived from the relativistic Lagrangian in Eq.~\eqref{eq:Lphi} (minimally coupled to gravity) by 
decomposing the ULDM field $\phi$ using \(\phi = [\psi \exp({-im_{\phi}t}) + \text{h.c.}]/\sqrt{2m_{\phi}}\) with $\psi$ a non-relativistic wavefunction, and neglecting higher-order terms in the limit $\ddot\psi \ll m_\phi\dot\psi \ll m_\phi^2\psi$. After a short derivation (see e.g. \cite{Guth:2014hsa,Eby:2018dat}), the resulting GPP equations are
\begin{align} \label{eq:GP}
    i \dot{\psi} &= \bigg(-\frac{\nabla^2}{2m_\phi} 
        + V_g(|\psi|^2) 
        - \frac{\lambda}{8m_{\phi}^2}|\psi|^2 \bigg) \psi\,,
                    \\ 
    \nabla^2 V_g &= 4\pi G m_{\phi}^2 |\psi|^2\,,
\end{align}
where \(V_g(|\psi|^2)\) is the self-gravitational potential.
When the density $|\psi|^2$ is relatively small, the third term in Eq.~\eqref{eq:GP} (which arises from self-interactions and is proportional to $\lambda$) can be neglected, and the boson star is understood as a stable balance of the other two forces.

As the mass of a boson star grows, its density also grows, and the attractive self-interaction of the field 
becomes stronger. Once it is of the same order as the other terms in Eq.~\eqref{eq:GP}, this interaction destabilizes the star, leading to gravitational collapse. This occurs once the boson star reaches a
critical value of the mass 
\(M_c \approx 10 \Mpl/\sqrt{\lambda}\) \cite{Chavanis:2011zi,Chavanis:2011zm}. The corresponding (minimum) radius of a boson star is \(R_c \approx 0.5 \Mpl\sqrt{\lambda}/m_{\phi}^2\).

\begin{figure}
    \centering
    \includegraphics[width = 1\linewidth]{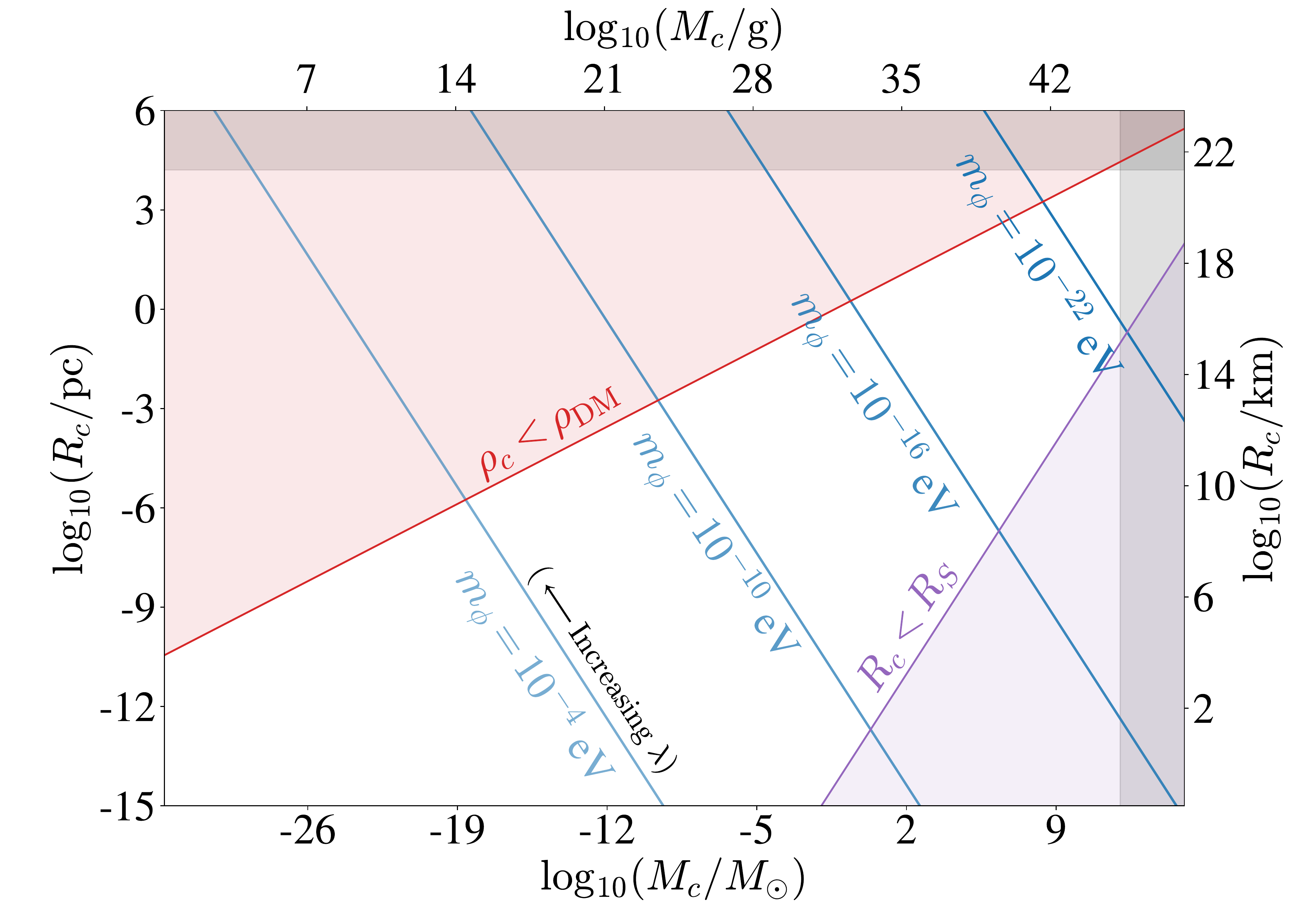}
    \caption{Mass-radius relation of scalar boson stars along the critical point in Eq.~\eqref{eq:RcMc} for different values of the scalar mass, \(m_{\phi}\) (blue lines, as labeled). The red shaded region is disfavored from the perspective of detection, as in this region, the density of 
    a boson star is smaller than the ambient DM density. Additionally, the two gray shaded regions signify where the boson stars either become larger (\(\gtrsim 16\) kpc), or more massive (\(\gtrsim 10^{12} M_{\odot}\)) than the Milky Way galaxy. The purple shaded region represents the Schwarzschild limit, where boson stars will not explode; see Section~\ref{sec:bosonstar} for details.}
    \label{fig:MassRadius}
\end{figure}

As a consequence of the above, at the critical point (just before it collapses), the radius and mass of a boson star are related as 
\begin{align} \label{eq:RcMc}
    R_c \approx \frac{5\Mpl^2}{M_c m_{\phi}^2}\,.
\end{align}
In Fig.~\ref{fig:MassRadius}, we show the mass-radius relationship of critical boson stars in Eq.~\eqref{eq:RcMc} for different choices of the \(m_{\phi}\) (blue lines, as labeled). The grey regions represent masses and radii of boson stars that are either more massive or larger than the Milky Way itself. The red region represents the region where the detection of bosenovae is disfavored because the density of the boson star  $\rho_c\simeq 3M_c/(4\pi R_c^3)<\rho_{\rm DM}$.

Importantly, the allowed mass-radius relationship for 
critical boson stars in Eq.~\eqref{eq:RcMc} was derived in the non-relativistic limit. 
As $m_\phi$ increases, the critical radius $R_c$ becomes smaller, eventually approaching the Schwarzschild radius of the boson star, \(R_S \equiv 2M_c/\Mpl^2\); at this point, the nonrelativistic calculation would suggest that the boson star forms a black hole, though in fact the calculation breaks down unless $R_c \ll R_S$.
Using Eq.~\eqref{eq:RcMc}, we can interpret this as a 
limitation on the critical mass as a function of \(m_{\phi}\), 
\begin{align}
    M_c \gtrsim \sqrt{\frac{5}{2}}\frac{\Mpl^2}{ m_{\phi}}\,,
\end{align}
or equivalently, a minimum value of \(\lambda\) 
\begin{align} \label{eq:lamdaBH}
    \lambda \gtrsim \lambda_{\rm BH} \equiv 40 \frac{ m_{\phi}^2}{\Mpl^2}\,,
\end{align}
for which our study is applicable. 

We illustrate this region of parameter space in purple color in both Figs.~\ref{fig:lambda_constraints} and \ref{fig:MassRadius}.
In the shaded range, the equation of state is dominated by general-relativistic corrections and the analysis above no longer applies. On the basis of previous work~\cite{Eby:2016cnq,Levkov:2016rkk,Helfer:2016ljl,Eby:2017xrr,Eby:2020ply}, we expect no bosenova to take place in this regime. 
Note that this applies, for example, to QCD axion stars for $m_\phi \lesssim 10^{-10}$\,eV (see green dotted line in Fig.~\ref{fig:lambda_constraints}).

\subsection{Bosenova signal}

\begin{figure*}[ht]
    \centering
    \includegraphics[width = 1\linewidth]{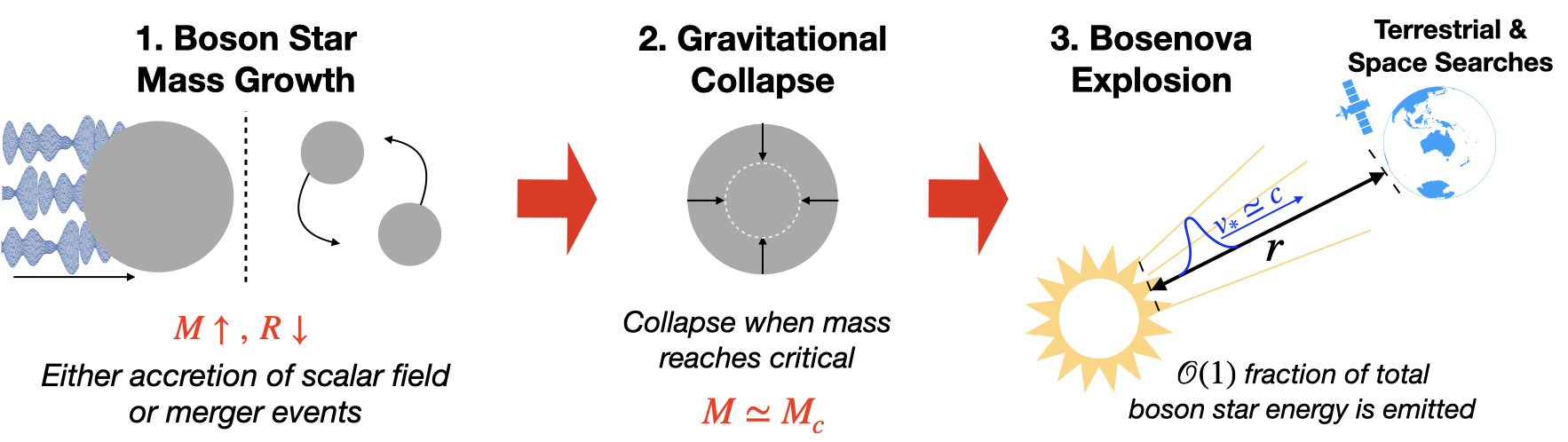}
    \caption{Illustration of the basic process leading to the signal analyzed in this work, which divides into three subprocesses: (1) The boson star mass $M$ grows by accretion or merger, until (2) $M$ approaches the critical value $M_c$, leading to spherical collapse. Finally, (3) in the final moments of collapse, the collapsing scalar field density grows large, exciting rapid emission of relativistic scalars in a \emph{bosenova} explosion. See text for further details.}
    \label{fig:bosenova}
\end{figure*}

As described in the Introduction, very massive boson stars (with mass approaching $M_c$) collapse gravitationally due to the attractive $\phi^4$ self-interaction in Eq.~\eqref{eq:Lphi}.
In the final stage of boson star collapse, relativistic number-changing processes in the core of the collapsing star are excited, and high-energy ULDM particles are rapidly emitted from the star \cite{Eby:2016cnq,Levkov:2016rkk}. This process is known as a \emph{bosenova}, by analogy to supernovae observed from collapse of heavy stars. The ``life-cycle'' of a boson star (including mass growth, collapse, and bosenova) is illustrated in Fig.~\ref{fig:bosenova}. 

A full-scale simulation of the collapse, including relativistic evolution and eventual bosenova, 
was conducted in \cite{Levkov:2016rkk}.
The authors found that 
the leading relativistic peak in the emission spectrum was centered near momentum of\footnote{This peak can be understood as arising from a leading $3\phi\to\phi$ number-changing interaction which is $\propto\lambda$; see e.g. \cite{Eby:2015hyx,Eby:2016cnq,Eby:2017azn}.} $\bar{k}\simeq 2.4m_\phi$, with width $\delta k \simeq m_\phi$. The total integrated energy emitted around this leading peak was of order
\begin{align} \label{eq:Epeak}
    \mathcal{E}_{\text{peak}} \approx 1020 \frac{m_{\phi}}{\lambda}\,,
\end{align}
where we translated the result of Ref.~\cite{Eby:2021ece} but substituted $\lambda = m_\phi^2/f_a^2$. In the simulation, the boson star collapsed and exploded $\mathcal{O}({\rm few})$ times before eventually settling to a stable, dilute configuration; each collapse produced an energy of order $\mathcal{E}_{\rm peak}$ on a timescale of order $\delta t_0 \simeq 400/m_\phi$, corresponding to an intrinsic burst `size' of $\delta x_0 \simeq 400/m_\phi$ (since the scalar velocity is $v_*\simeq 1$). This work focuses on the signal from a single explosion, and is in this sense conservative, as the signal should only be enhanced by additional subsequent explosions. 

Note that the authors of~\cite{Levkov:2016rkk} analyzed the emission spectrum for axion-like fields with a chiral potential, i.e. a QCD axion (see e.g. \cite{GrillidiCortona:2015jxo}).
The leading term in this potential has the same form as we consider in Eq.~\eqref{eq:Lphi}.
Although we are considering a different set of physical models, we will employ the output of these simulations as a characteristic behavior for the present study. Indeed, the authors of~\cite{Levkov:2016rkk} indicate that 
in their simulations, they did not observe a strong dependence of the emission spectrum 
on the specific form of the potential, and since we are considering only the leading relativistic peak (which should be dominated by the $\phi^4$ term), we expect the difference to be small at leading order.

The precise form of the potential, including higher-order terms, could affect detection prospects.
It is an open question, detailed study of which we leave for future work, whether differences in the emission spectrum arising from distinct self-interaction potentials can be distinguished experimentally. In the event of a detection, such details could contain important information about the underlying scalar field theory, including its UV completion, which could be challenging to probe otherwise.

Bosenovae eject bursts of relativistic ULDM bosons in an approximately spherical shell, which can eventually reach Earth. Under the assumption that the thickness bosenova shell \(\delta x\) is much smaller than the distance from the explosion to Earth, \(\delta x \ll r\), the energy density in a bosenova shell is 
\begin{align}
    \rho_{*} \approx \frac{\mathcal{E}}{4\pi r^2 \delta x}
\end{align}

\noindent where \(\mathcal{E}\) is the total energy emitted in the burst. 
If the emitted scalar waves do not spread in flight, then the duration of the burst is merely the intrinsic duration from the source, $\delta x\to \delta x_0$. 
Using this result and substituting $\mathcal{E}=\mathcal{E}_{\rm peak}$ from Eq.~\eqref{eq:Epeak}, one finds 
\begin{align}
    \rho_{*,0} \simeq  4\cdot 10^{8} \rho_{\rm DM} \bigg(\frac{m_{\phi}}{10^{-15} ~\text{eV}}\bigg)^2\bigg(\frac{10^{-80}}{\lambda}\bigg) \bigg(\frac{\rm pc}{r}\bigg)^2
    \label{eq:density}
\end{align} 
in the limit of minimal wave spreading. 
However, as the shell propagates to Earth, the wave naturally spreads in flight. This further dilutes the energy density by the factor \cite{Eby:2021ece}
\begin{align}
    \frac{\delta x_0}{\delta x} \simeq \bigg(\frac{\xi^2 q(q^2 + 1)}{1.5\times 10^{8}}\bigg) \bigg(\frac{10^{-15}~\text{eV}}{m_{\phi}}\bigg)\bigg(\frac{\text{pc}}{r}\bigg) \,.
\end{align}
where \(\xi \equiv m_\phi \delta x_0 \approx 400\), and \(q \equiv \bar{k}/m_\phi \approx 2.4\), and we took $\delta k \simeq m_\phi$. 
Therefore, at the detector, the energy density takes the form
\begin{align}
\label{eq:density_wavespread}
    \frac{\rho_*}{\rho_{\rm DM}} \simeq \frac{\rho_{*,0}}{\rho_{\rm DM}}
    \frac{\delta x_0}{\delta x} \simeq 3
    \bigg(\frac{m_{\phi}}{10^{-15}~\text{eV}}\bigg)
    \bigg(\frac{10^{-80}}{\lambda}\bigg) \bigg(\frac{\rm pc}{r}\bigg)^3\,.
\end{align}
We conclude that the energy density in the burst at the position of Earth can be much larger than the ambient DM density $\rho_{\rm DM} \simeq 0.4\,{\rm GeV/cm}^3$.

On the other hand, the oscillations of the scalar waves are very incoherent at the source (with $\mathcal{O}(m_\phi)$ spread in momentum), 
compared to the high-quality oscillations of the ambient DM (which has a momentum spread $\propto m_\phi v_{\rm DM}^2 \simeq 10^{-6}m_\phi$).
However, as pointed out in \cite{Eby:2021ece}, the spreading of the waves leads to increased effective coherence in the detector.
This can be understood by observing that the different momentum modes travel at different velocities, e.g. with the fastest modes arriving first, so at any given time the energy deposition in the detector is remarkably coherent. It was found that the \emph{effective} coherence time of a relativistic burst at the detector scaled as
\begin{equation} \label{eq:taustar}
    \tau_* \simeq \frac{2\pi r}{q^3 \xi} 
    \simeq 4\cdot10^{-3}\,{\rm year}\left(\frac{r}{{\rm pc}}\right)\,,
\end{equation}
which approaches e.g. $1$ year for $r \simeq 300$ pc. The duration of the burst is also extended, as
\begin{equation} \label{eq:tburst}
    \delta t \simeq \frac{\delta k}{m_\phi} \frac{r}{q^2\sqrt{q^2+1}}
    \simeq 0.2\,{\rm year} \left(\frac{r}{{\rm pc}}\right)\,,
\end{equation}
where as before we took $\delta k \simeq m_\phi$. Note that this large momentum spread implies that broadband (rather than resonant) searches for ULDM are optimal for bosenova searches. 

Taking the above effects into account, we can characterize the sensitivity of a search for relativistic axions, $d_{i,*}$, relative to a DM search sensitivity, $d_{i}$, at the same frequency by the ratio \cite{Eby:2021ece}
\begin{align}
    \label{eq:gratio}
    \frac{d^{(1)}_{i,*}}{d^{(1)}_{i}} \sim \sqrt{\frac{\rho_{\text{DM}}}{\rho_*}} \frac{t_{\text{int}}^{1/4} \text{min}\bigg(\tau_{\rm{DM}}^{1/4},~t_{\text{int}}^{1/4}\bigg)}{\text{min}\bigg((\delta t)^{1/4},~t_{\text{int}}^{1/4}\bigg)~\text{min}\bigg(\tau_{*}^{1/4},~t_{\text{int}}^{1/4}\bigg)}~,
\end{align}
where \(t_{\text{int}}\) is the integration time of the experiment. We observe for example that for long burst duration ($\delta t \gtrsim t_{\rm int}$) and long effective coherence time ($\tau_* \to \tau_{\rm DM}$), the sensitivity ratio is determined solely by the energy density in the burst $\rho_*$ relative to the background DM density $\rho_{\rm DM}$. We will discuss the sensitivity $d_{i,{\rm DM}}^{(1)}$ of current and future experimental searches in Section~\ref{sec:detection}.

\subsection{Bosenova rate}

The event rate of bosenovae is sensitive to model-dependent cosmological and astrophysical assumptions, and hence is complicated with non-trivial estimates. To a first approximation, the rate depends on the intrinsic formation rate and distribution of boson stars at high redshift, the accretion \cite{Eggemeier:2019jsu,Chen:2020cef,Chan:2022bkz,Dmitriev:2023ipv} and merger \cite{Eby:2017xaw,Hertzberg:2020dbk,Du:2023jxh} rate of boson stars, and details of the collapse and bosenova processes \cite{Eby:2016cnq,Levkov:2016rkk,Eby:2017xrr,Levkov:2020txo}, detailed studies of which are beyond the scope of our work focusing on detection.
In this subsection we provide crude estimates, and additional discussion of the relevant assumptions can be found in Ref.~\cite{Eby:2021ece}. 

For our purposes, we employ the simplified ansatz of a fixed boson star mass $M=M_c$ and assume a homogeneous distribution around the galaxy, taking the fraction of DM in boson stars to be $f_{\rm DM} \leq 1$. In this case, 
the average distance \(\bar{r}\) between boson stars is
\begin{align}
    \bar{r} &\approx \bigg(\frac{3M_c}{4\pi f_{\text{DM}}\rho_{\text{DM}}}\bigg)^{1/3}
    \approx \frac{30\,{\rm pc}}{f_{\rm DM}^{1/3}}
    \left(\frac{10^{-80}}{\lambda}\right)^{1/6}\,.
\end{align}
This estimate would imply a total number of boson stars
\begin{equation} \label{eq:Nbs}
 N_{\rm bs} \sim 4\cdot 10^7 f_{\rm DM}
 \left(\frac{\lambda}{10^{-80}}\right)^{1/2}  
 \left(\frac{r}{10\,{\rm kpc}}\right)^3
\end{equation}
in a galaxy of size $\sim 10$ kpc (e.g. the Milky Way). 

As mentioned above, the average rate of bosenova signals is complicated with significant uncertainties (see~\cite{Eby:2021ece,Escudero:2023vgv} for some discussion). Let us consider an approximate simplified example. Suppose one boson star collapse occurs on average every $T\sim$ Gyr within a given galaxy. Then using the estimate in Eq.~\eqref{eq:Nbs}, the rate of bosenovae within a distance $r$ of Earth would be of order 
\begin{equation}
  \Gamma = \frac{N_{\rm bs}}{T}
        \sim 0.04\,{\rm year}^{-1}
        \left(\frac{\lambda}{10^{-80}}\right)^{1/2}\left(\frac{r}{10\,{\rm kpc}}\right)^3\,.
\end{equation}
Therefore, this would imply one bosenova every $25$ years in the Milky Way. For comparison, the supernova rate in the Milky Way is estimated at roughly $1$ per $50$ years~\cite{Rozwadowska:2020nab}.
We postpone further discussion and improved estimates of the bosenova rates for future work.

\section{Methods of Detection}
\label{sec:detection}

A variety of experimental setups and technology, especially those traditionally focusing on detecting scalar DM (see e.g.~\cite{Antypas:2022asj}) and variation of fundamental constants, constitute excellent laboratories for probing bursts of propagating relativistic ULDM particles. The sensitivity to couplings is further enhanced when the density of the scalar particles originating from a bosenova (or other) burst at the detector is larger than the ambient scalar particle DM density, in accordance with Eq.~\eqref{eq:gratio}.

Various technologies for the detection of scalar DM have recently been reviewed in \cite{Antypas:2022asj}. These include atomic, molecular, and nuclear clocks, as well as other spectroscopy experiments, optical cavities, atom interferometers, optical interferometers, torsion balances, mechanical resonators, and others. While here were are detecting the transient rather than continuous signal due to halo DM, the detection principles remain the same and we only briefly review the sensitivities of relevant current and future detectors below. 

\subsection{Atomic clocks}
The coupling of the scalar DM to the SM leads to oscillations of fundamental constants \cite{Arvanitaki:2014faa}, such as the fine-structure constant $\alpha$, proton-to-electron mass ratio $\mu=m_p/m_e$, and  $X_q=m_q/\Lambda_\textrm{QCD}$ \cite{Flambaum:2006quarks}, where $m_q$ is the average light quark mass, and $\Lambda_{\rm{QCD}}$ is the QCD scale. As atomic, molecular, and nuclear energy levels depends on values of fundamental constants, this effect leads to the variation of such energies, as well as the clock frequencies.  Different types of clocks are sensitive to different fundamental constants. Moreover, clocks based on different transitions could have significantly different sensitivities; therefore,  one observes a ratio of clock frequencies over time and extract the signal via the discrete Fourier transform or power spectrum of the data \cite{Arvanitaki:2014faa,Hees:2016gop,2022PhRvL.129x1301K,Beloy:2020tgz,Sherrill:2023zah,Filzinger:2023zrs}. The signal will persist for the duration of the burst.  
The bosenova burst would be detectable with various quantum clocks for a wide range of DM masses (being most sensitive for $m  \lesssim 10^{-13}$ eV) and interaction strengths.

\paragraph{\textbf{Optical atomic clocks}} 
The ratios of optical clocks frequencies \cite{ludlow_optical_2015}, i.e., based on transitions between different electronic levels (frequencies of $0.4-1.1\times10^{15}$~Hz)
are sensitive to photon \cite{Arvanitaki:2014faa} and hadronic sector \cite{Banerjee:2023bjc} couplings. 
The frequency of the optical atomic clock  can be expressed as \cite{SafBudDem18}
\begin{equation}
\nu \sim c R_{\infty} A F(\alpha),
\label{optical}
\end{equation}
where $c$ is the speed of light, $A$ is a numerical factor depending on an atom, $F(\alpha)$
depends upon the particular transition, and $R_{\infty}$ is the Rydberg constant.
The sensitivity of optical atomic clocks to the variation of $\alpha$ is  parameterized by dimensionless sensitivity factors $K$ that can be computed from first principles with high precision \cite{FlaDzu09} and can be either positive or negative. Ultimate accuracy in the ability to detect the variation of the fundamental constant, and, therefore, ultralight scalar DM, depends on difference in the sensitivity factors between two clocks and the achievable  fractional accuracy of the ratio of frequencies $\nu$:
\begin{equation}
\frac{\partial}{\partial t} \textrm{ln}\frac{\nu_2}{\nu_1}=(K_2-K_1)\frac{1}{\alpha}\frac{\partial \alpha}{\partial t}\,,
\label{K}
\end{equation}
where indices 1 and 2 refer to clocks 1 and 2, respectively.
The Yb$^+$ clock based on an electric octupole (E3) $4f^{14}6s\,^2S_{1/2} \leftrightarrow 4f^{13}6s^2\,^2F_{7/2} $ transition has the largest (in magnitude) sensitivity factor $K=-6$ \cite{FlaDzu09} among all presently operating clocks. The Yb$^+$ ions also supports another clock transition based on the electric quadrupole (E2) $4f^{14}6s\,^2S_{1/2} \leftrightarrow 4f^{14}5d\,^2D_{3/2} $  transition with $K=1$ and $^{171}$Yb$^+$ E3/E2 clock-comparison pair presently provides the best limit for scalar DM for the lightest masses, with the experiment carried out by the PTB team \cite{Filzinger:2023zrs}.

Optical clocks based on highly charged ions (HCIs) \cite{kozlov_highly_2018} will have much larger sensitivities to the variation of $\alpha$ than presently-operating optical clocks. For example, a Cf$^{15+}$/Cf$^{17+}$ pair has $K_2-K_1 > 100$ \cite{2020Cf}. Development of HCI clocks is progressing rapidly, with a recent demonstration of Ar$^{13+}$ clock with $2\times 10^{-17}$ uncertainty \cite{2022HCI}.

Recently, it was shown that coupling of ultralight scalar DM to quarks and gluons would lead to an oscillation of the nuclear charge radius detectable with optical atomic clocks \cite{Banerjee:2023bjc,flambaum2023variation}, and their comparisons can be used to investigate DM-nuclear couplings, which were previously only accessible with other platforms. 

The total electronic energy $E_\textrm{tot}$ of an atomic state contains the energies associated with the finite nucleus mass (mass shift, MS) and the non-zero nuclear charge radius $r_N$ (field shift, FS), which dominated for heavy atoms and provides the better sensitivity.  The field shift can be parameterized as
$$E_{\rm FS} \simeq K_{\rm FS} \left<r_N^2\right>,$$
where  $K_{\rm FS}$ is the field-shift constant leading to oscillation of the atomic energy due to the oscillation of the  $r_N$ caused by the coupling of nuclear sector to DM \cite{Banerjee:2023bjc}.  

Therefore, measuring the ratio of two clock frequencies  $\nu_2$ and $\nu_1$ of heavy atoms enables the detection of ultralight DM that will cause the oscillation of $r_N$:
\begin{align}
\frac{\Delta (\nu_2/\nu_1)}{(\nu_2/\nu_1)} = K_{2,1} \frac{\Delta \left<r_N^2\right>}{\left<r_N^2\right>}.\label{eq:d_nu_lim}
\end{align}
 We note that the sensitivity  $K_{2,1}$ of the clock pair to DM is different from the sensitivity to $\alpha$ and is determined by the field shift constants of the clock transitions: 
$$
    K_{2,1} \equiv \frac{K^{\nu_2}_{\rm FS} \,\langle r_N^2 \rangle}{\nu_2}
   -\frac{K^{\nu_1}_{\rm FS}\,\langle r_N^2 \rangle}{\nu_1} \,.
$$
Corresponding limits on DM to coupling to gluons $d_g$ and quarks $d_{\hat m}$ are obtained as 
$$
\frac{\Delta (\nu_2/\nu_1)}{(\nu_2/\nu_1)} = K_{2,1} \Big[ (c_1+c_2) d_g + c_2 d_{\hat m}\Big] \,\sqrt{4\pi}\frac{\sqrt{2 \rho_{\rm DM}}}{m_\phi\,\Mpl}\,,
 $$
where $c_1$ and $c_2$ are of order unity (see Refs.~\cite{Banerjee:2023bjc,flambaum2023variation} \footnote{The coefficients $c_1$ and $c_2$ were originally labelled as $\alpha$ and $\beta$ (respectively) in \cite{Banerjee:2023bjc}. We have changed them here to avoid confusion with $\alpha$, $\alpha_s$, and $\beta(g_s)$.}). 
Interestingly, $^{171}$Yb$^+$ E2/E3 clock pair has high sensitivity to this effect as well --- upper states in these two clock transitions have significantly different electronic structure resulting in field shifts constants that differ in sign.  Yb is also quite heavy with $Z=70$ and near-future experiments will allow improvement in a wide mass range \cite{Banerjee:2023bjc}. 
 
\paragraph{\textbf{Microwave clocks}} Microwave clocks are based  on transitions between hyperfine substates of the ground state of the atom (frequencies of a few GHz).  The corresponding frequency of such with  transitions  can
be expressed as \cite{SafBudDem18}
\begin{equation}
\nu_\textrm{hfs} \sim c R_{\infty} A_\textrm{hfs} \times g_i \times \frac{m_\textrm{e}}{m_\textrm{p}} \times \alpha^2 F_\textrm{hfs}(\alpha),
\label{eq2}
\end{equation}
where $A_\textrm{hfs}$ is a numerical quantity depending on a particular atom
 and
$F_\textrm{hfs}(\alpha)$ is specific to each hyperfine transition.
The dimensionless quantity $g_i=\mu_i/(I \mu_\textrm{N})$   is the  nuclear $g$-factor, where $\mu_i$ is the nuclear magnetic moment, $I$ is a nuclear spin, and 
$\mu_\textrm{N}=\frac{e\hbar}{2m_\textrm{p}}$ is the nuclear magneton. The  variation of $g$-factors is commonly reduced to the variation of $X_q=m_q/\Lambda_{\textrm{QCD}}$ enabling sensitivity to DM-SM coupling to gluons $d_g$ and quarks $d_{\hat m}$.

Comparing formulas given by Eq.~(\ref{optical}) and Eq.~(\ref{eq2}), we see  that ratio of microwave to optical clocks is sensitive to the various of $\alpha$, $\mu$, and $X_q$ providing sensitivity to all linear couplings discussed in this work. Microwave Rb to Cs frequency clock ratio is sensitive to variation of $\alpha$ and 
$X_q$ \cite{SafBudDem18}:
  \begin{equation}
 \frac{\nu_{\textrm{Cs}}}{\nu_{\textrm{Rb}}}= \frac{g_{\textrm{Cs}}}{g_{\textrm{Rb}}}  \alpha^{0.49},
  \end{equation}
see 
Ref.~\cite{Flambaum:2006quarks,2019MSreview} for extraction of sensitivity to nuclear sector from the $g$-factors. Therefore, microwave clocks can probe all of the scalar DM couplings  discussed here, but have reached their technical accuracy limit of $10^{-16}$~\cite{2018CsClock} , two orders of magnitude below the optical clocks. Rb/Cs clock-comparison limits are reported in \cite{Hees:2016gop}.

\subsection{Molecular and nuclear clocks}
Several new types of clocks are being developed (see reviews \cite{Antypas:2022asj,2019MSreview}), based on molecules and molecular ions \cite{2021Hanneke,ZelevinskyKondovNPhys19_MolecularClock}, and the $^{229}$Th nucleus \cite{Peik2021}.

 Molecular clocks provide enhanced sensitivity to 
$\mu$-variation and will allow a significantly improved sensitivity to electron couplings as well. For example, the linear triatomic molecule SrOH possesses a low-lying pair of near-degenerate vibrational states leading to a large sensitivity to changes in $\mu$ and a high degree of control over systematic errors~\cite{SrOH2021}.

The design of a high precision optical clocks requires ability to construct a laser operating at the wavelength of the clock transition, which precludes using 
 nuclear energy levels as their transition frequencies are generally outside of the laser-accessible range by many orders of magnitude. However, there is (so far) a single known exception, 
a  nuclear transition that occurs between the long-lived (isomeric) first excited state of the $^{229}$Th and the corresponding nuclear ground state, with a laser-accessible wavelength near 149\,nm. 

In 2023, the first observation of the radiative decay of the $^{229}$Th nuclear clock isomer was reported \cite{Kraemer_2023}, and 
the transition energy was  measured to be 8.338(24) eV, corresponding to the photon vacuum wavelength of the isomer's decay of 148.71(42) nm. The nuclear clock can be operated with a single Th$^{3+}$ trapped ion, much like a single ion atomic clock, except that it excites a nuclear rather than atomic transition.  Laser cooling of Th$^{3+}$   has already been demonstrated.  An alternate solid-state scheme has also been suggested which can not be implemented with atomic clocks (see the review \cite{Peik2021} and references therein).

Nuclear clock is expected to have several orders of magnitude larger sensitivities to both the variation of $\alpha$ and $X_q$, giving a unique opportunity to drastically enhance scalar ULDM searches, since the projected accuracy of nuclear trapped ion clocks is 10$^{-19}$ \cite{2012PhRvL.108l0802C}.
Flambaum \cite{FlaTh06} suggested that the anomalously small transition energy of the $^{229\textrm{m}}$Th isomer is the result of a nearly perfect cancellation of a change in Coulomb energy $$\Delta E_C=E^{\rm{m}}_C-E_C \sim -1~\textrm{MeV}$$ by opposite and nearly equal changes of the nuclear energy through the strong interaction; this is why both sensitivity to photon and nuclear couplings are similarly enhanced.  
This difference in the Coulomb energy and corresponding  sensitivity to variation of $\alpha$ can be estimated from the measured differences in the charge radii and quadrupole moments $Q_0$ between the ground state and the isomer \cite{BerDzuFla09,ThiOkhGlo17}: 
\begin{eqnarray}
\Delta E_C\approx&-&485 ~\textrm{MeV}~\left[\frac{\langle r^2_{229\textrm{m}} \rangle} {\langle r^2_{229} \rangle} -1 \right]  \\
&+& 11.6~ \textrm{MeV}~ \left[Q^{\textrm{m}}_0/Q_0-1 \right]=-0.29(43)~ \textrm{MeV},\nonumber
\end{eqnarray}
which is limited by the accuracy in the $Q^{\textrm{m}}_0/Q_0$ value, but planned to be improved with a better ion trap \cite{Peik2021}. The sensitivity value of $K = -0.9(3) \times 10^4$ for $\alpha$-sensitivity has been evaluated based on additional nuclear modeling  that includes a relation between the change of the charge radius and that of the electric quadrupole moment \cite{Fla2020}.
Using this value, the sensitivity to the variation of $\alpha$ and DM is determined the same way as for the atomic clock pair given by Eq.~(\ref{optical}). A Th nuclear clock can be compared to any other atomic clock or a cavity. The sensitivity for the DM couplings of the hadronic sector is also $\mathcal{O}(K)$; see \cite{FlaTh06}.
The plan for the development of a nuclear clock have been discussed in detail in  \cite{Peik2021} and rapid progress is expected after a recent new measurement of the transition wavelength \cite{Kraemer_2023}.

\subsection{Optical cavities}
 Variations of $\alpha$ and particle masses also alter the geometric sizes of solid objects, scaling as $L \propto a_\textrm{B}$, where $a_\textrm{B} = 1/(m_e \alpha)$ is the atomic Bohr radius \cite{Stadnik1,Stadnik2} in the non-relativistic limit. 
 When sound-wave propagation through the solid occurs sufficiently fast for a solid to fully respond to changes in the fundamental constants, the size of a solid body changes according to \cite{Antypas:2022asj}: 
\begin{equation}
    \label{adiabatic_solid_length_change}
    \frac{\delta L}{L} \approx \frac{\delta a_\textrm{B}}{a_\textrm{B}} = - \frac{\delta \alpha}{\alpha} - \frac{\delta m_e}{m_e} \, . 
\end{equation}
 leading to the sensitivity to both photon and electron couplings. 

Therefore, the cavity reference frequency $\nu_\textrm{cavity} \propto 1/L$ can respond to changes in the fundamental constants, as~\cite{Stadnik1,Stadnik2,Antypas:2022asj}
\begin{equation}
    \label{free_cavity_length_changes}
\frac{\delta \nu_\textrm{cavity}}{\nu_\textrm{cavity}} = - \frac{\delta L}{L} \approx -\frac{\delta a_\textrm{B}}{a_\textrm{B}} = \frac{\delta \alpha}{\alpha} + \frac{\delta m_e}{m_e}
\end{equation}
for the cavity whose length depends on the length of the solid spacer between the mirrors, 
enabling DM searches.
One can compare the cavity reference frequency to the atomic frequency 
or to another cavity. We note that the atomic clock design includes an optical cavity, naturally supporting clock-cavity comparisons, as carried out in~\cite{Kennedy:2020bac}. 

 The size changes of the solid are enhanced if the oscillation frequency of the fundamental constants matches the frequency of a fundamental vibrational mode of the solid \cite{Antypas:2022asj}. Various optical cavities can be used for ultralight scalar particle detection, with comparisons of atom-cavity \cite{Campbell:2020fvq} and cavity-cavity \cite{Oswald:2021vtc,Tretiak:2022ndx} comparisons setting limits on DM-SM couplings. Such experiments are naturally sensitive to DM mass ranges higher than that of clocks, complementing clock-clock comparisons as well as providing  limits on additional couplings. 

\subsection{Atom interferometers}
In atom interferometry, laser pulses are used to coherently split, redirect, and recombine matter waves through stimulated
absorption and emission of photons, driving transitions between the ground and a long-lived excited atomic state \cite{SnowmassAI}. We note that $^1S_0-^3P_0$ narrow transition in Sr is used for both ultraprecise atomic clocks and atom interferometry schemes for future gravitational wave detection and DM searches \cite{2021QS&T....6d4003A}. 

Atom interferometers are  sensitive to the oscillations of fundamental constants  as well as DM-induced accelerations in dual-species interferometers, for example $^{88}$Sr and $^{87}$Sr.  Such experiments compare the phase accumulated by delocalized atom clouds with DM affecting the atomic energies or exerting a force on the atom clouds, respectively.

Large-scale atom interferometers are required to enhance the sensitivity to DM. Large-scale 100-meter prototypes are currently being built \cite{2021QS&T....6d4003A}. Space-based atom  interferometers (see  AEDGE proposal \cite{Badurina:2021rgt} ) will be highly sensitive to scalar and vector DM.

 \subsection{Other detectors}
Other spectroscopy DM limits include atomic Dy
   \cite{VanTilburg:2015oza} and molecular iodine
   \cite{Oswald:2021vtc} experiments.
  
 Ultralight DM may also exert time-varying forces on test bodies in optical inteferometers, but such forces are suppressed by the smallness of the electromagnetic and electron-mass contributions to the overall mass of a test body for scalar DM (see \cite{Antypas:2022asj} for a brief review). 
 
 Torsion balances sense differential forces on macroscopic test masses. Although they were designed to test the equivalence principle (EP) \cite{Hees:2018fpg}, one can derive limits to scalar DM assuming that these particles create a differential force. MICROSCOPE space mission tested the EP in orbit using electrostatic accelerometers onboard a drag-free microsatellite \cite{Berge:2017ovy}. We note that all limits derived from EP tests do not assume DM halo density, so these limits will not be affected by the enhanced density. 
 
Various proposals recently reviewed in \cite{Antypas:2022asj,2021Carney}  use mechanical resonators  spanning a range of frequencies from 1\,Hz to 1\,GHz, corresponding to a UDM mass from $10^{-14}$\,eV to $10^{-5}$\,eV. These are naturally narrowband in DM mass. 

\subsection{Networks of detectors on Earth and in space}
One can use all of these experiments to set the limits on the bosenova bursts. The potential for significantly-increased density of particles compared to halo DM leads to an increased chance of detection for all density-dependent experiments. 
One significant factor to consider is whether the experiment can reach its ultimate sensitivity during the burst, which we assume in all limits. Due to vast differences in integration time to full sensitivity for different detection methods, we assume it can be achieved during the burst time in the present manuscript. 

A network of detectors allows for better differentiation of burst signals from the local noise sources as well as improved precision \cite{Antypas:2022asj}. Networks of detectors in space provide particularly-interesting opportunities due to the large possible distances between network nodes. For example, GPS has been used to provide limits on transient DM effects \cite{2017NatCo...8.1195R}. A potentially-tantalizing opportunity with the space network is the observation of the onset of the burst signal propagating through the network nodes and potential localization of the signal. Further work is needed for different detector types to explore these opportunities.
Detecting the onset of relativistic bursts as opposed to non-relativistic transients is a challenging prospect, but quantum technologies are improving rapidly, leading to significant improvements in both accuracy and stability. For example, the stability of optical lattice clocks has improved drastically in recent years, allowing us to reach 10$^{-18}$ uncertainties within a few minutes \cite{2022Natur.602..420B}, with further orders of magnitude improvements possible.  

Finally, note that a number of fundamental physics and DM studies with high-precision optical clocks in space have recently been proposed \cite{2023NA,FOCOS,OACESS}.

\begin{figure*}[ht]
    \centering
    \hspace{-2.0cm}
    \includegraphics[width = 0.9\paperwidth]{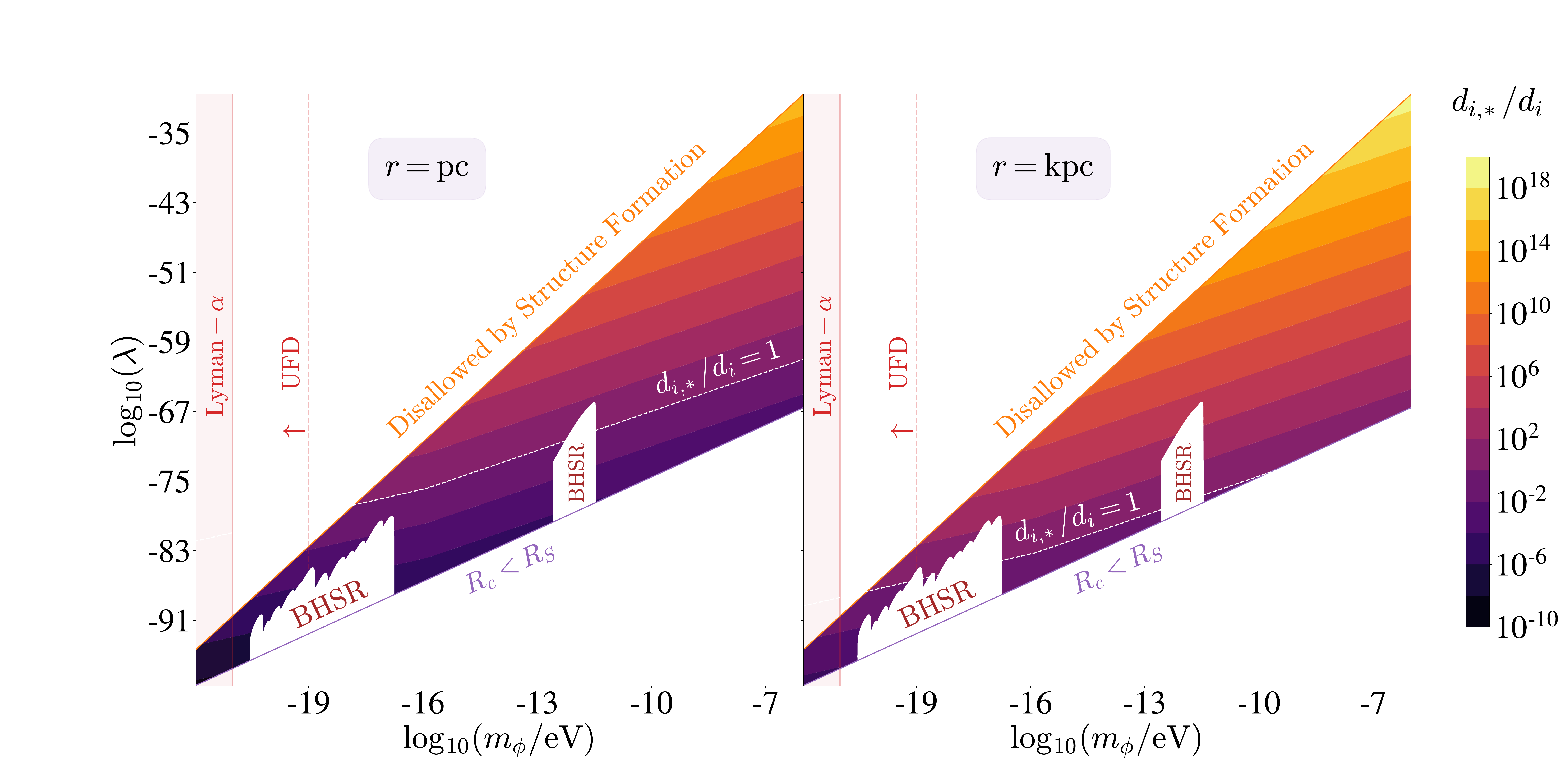}
    \caption{The ratio of sensitivity to a given coupling for a burst search $d_{i,*}$ relative to a DM search $d_i$, for the allowed parameter space in \(m_{\phi}\) vs \(\lambda\); the white dotted line corresponds to equal sensitivity to bursts and DM. We illustrate two choices of distance to the bosenova: \(r = \rm{pc}\) (left) and \(r = \rm{kpc}\) (right). See Section~\ref{ssec:lambdaconstraints} and Fig.~\ref{fig:lambda_constraints} for discussion of constraints.}
    \label{fig:gratio_contour}
\end{figure*}

\section{Results}
\label{sec:results}

We find that bursts of relativistic scalars offer a discovery reach orders of magnitude better than from background DM over a range of bosenova distances $r$, scalar masses $m_\phi$, and self-interaction couplings $\lambda$. In Fig.~\ref{fig:gratio_contour}, we show contours of the potential reach of the coupling ratio, \(d_{i,*}^{(1)}/d_{i}^{(1)}\) with $i=e,m_e,g$, over the \(m_{\phi}\)-\(\lambda\) plane. We calculate the coupling ratio using Eq.~(\ref{eq:gratio}), with burst density given by Eq.~\eqref{eq:density_wavespread} and the burst timescales in Eqs.~(\ref{eq:taustar}-\ref{eq:tburst}). Since the distance to the bosenova is a free parameter (for our purposes), we take two benchmarks of \(r = 1\) pc (left) and \(r = 1\) kpc (right) to capture a range of possible galactic sources. This coupling ratio is independent of the specifics of the experiment in question, and therefore provides a blueprint for 
the most interesting parameter space 
for current and future experiments. As in Fig.~\ref{fig:lambda_constraints}, the relevant parameter space is limited by Lyman-$\alpha$~\cite{Irsic:2017yje,Rogers:2020ltq} and UFDs~\cite{Dalal:2022rmp} and structure formation considerations (see Section~\ref{ssec:LSS}), as well as at small $\lambda$ by the limitations of the non-relativistic analysis (see Eq.~\eqref{eq:lamdaBH}).

The dotted white line in Fig.~\ref{fig:gratio_contour} represents where \(d_{i,*}^{(1)}/d_{i}^{(1)} = 1\), i.e. the coupling that can be probed by detecting a bosenova at a distance $r$ is the same as that of background DM. Modulo the timing factors of Eq.~\eqref{eq:gratio}, this in essence says that the density of scalars from the burst is the same as the background DM. Regions below this 
contour represent \(d_{i,*}^{(1)}/d_{i}^{(1)} < 1 \), which means that the coupling that can be probed in those regions of the parameter space are smaller than those that can be probed by signals from background DM. In other words, below this line 
the bosenovae are most advantageous. 

Fig.~\ref{fig:gratio_contour} can be used to translate an experiment's sensitivity to background DM into a sensitivity to a bosenova by multiplying the experimental coupling reach by the coupling ratio \(d_{i,*}/d_{i}\). The recipe to translate a DM limit to a burst limit in a given experiment is as follows: for a given mass range $m_{\phi}$ that the experiment is sensitive to, choose a value of $\lambda$ that is allowed in that mass range. Then the coupling reach can be determined by multiplying the 
DM limit on $d_i^{(1)}$ by the value of the $d_{i,*}^{(1)}/d_i^{(1)}$ over that $\lambda$ slice. For instance, if an experiment was sensitive to DM masses of $10^{-19} - 10^{-16}$ eV, a viable benchmark for $\lambda$ could be, e.g., $10^{-85}$. For a bosenova with these parameters, the coupling ratio $d_{i,*}^{(1)}/d_i^{(1)}$ is $\Ocal(10^{-3})$ for a distance of $r =1$ pc. This means that, in the presence of a bosenova, the sensitivity for the burst signal covers three additional orders of magnitude 
compared to the background DM limit.

\begin{figure*}
    \centering
    \makebox[ \paperwidth]{\hspace{-3.5cm}\includegraphics[width = .6\linewidth]{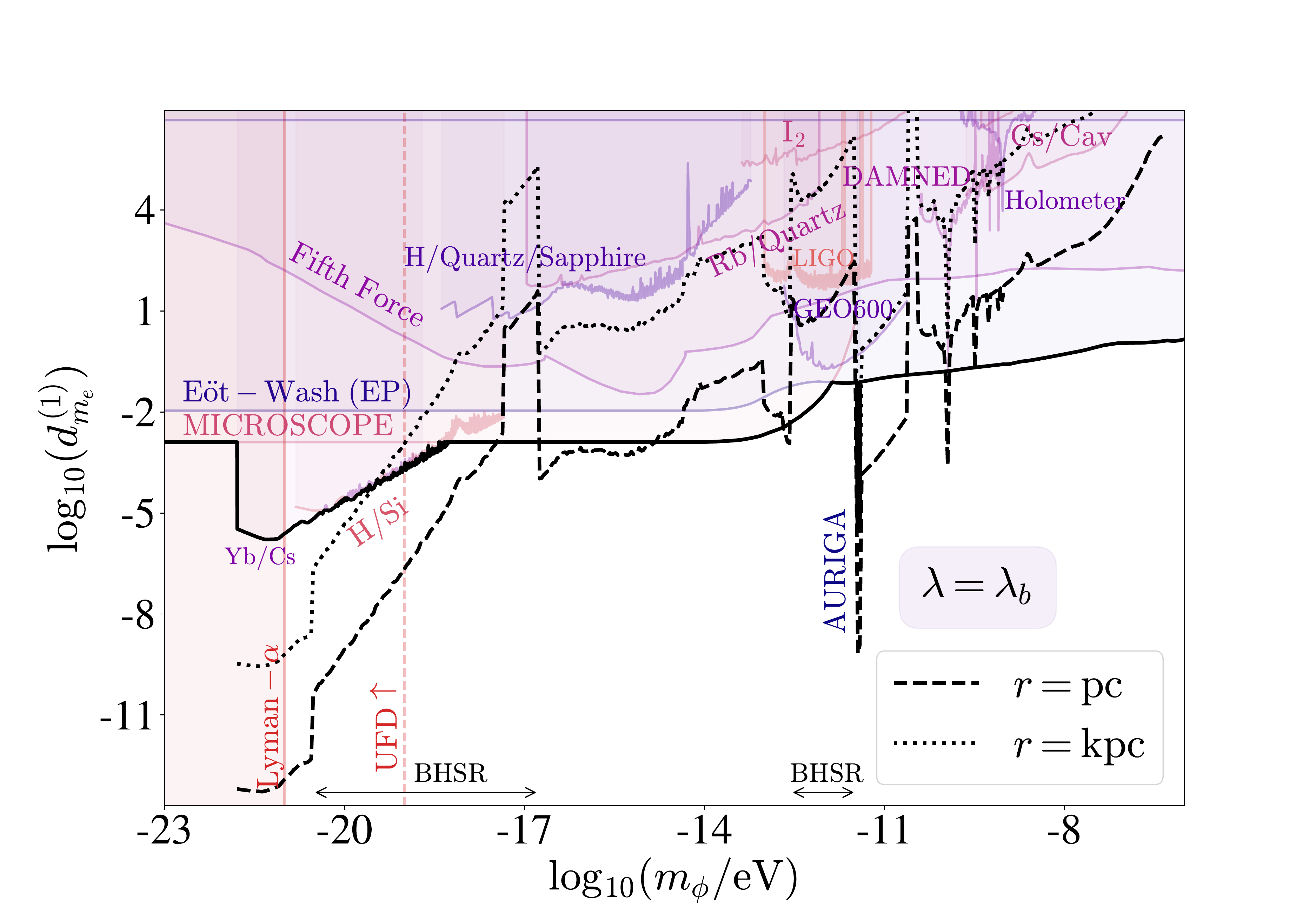}
    \hspace{-0.9cm}
    \includegraphics[width = .6\linewidth]{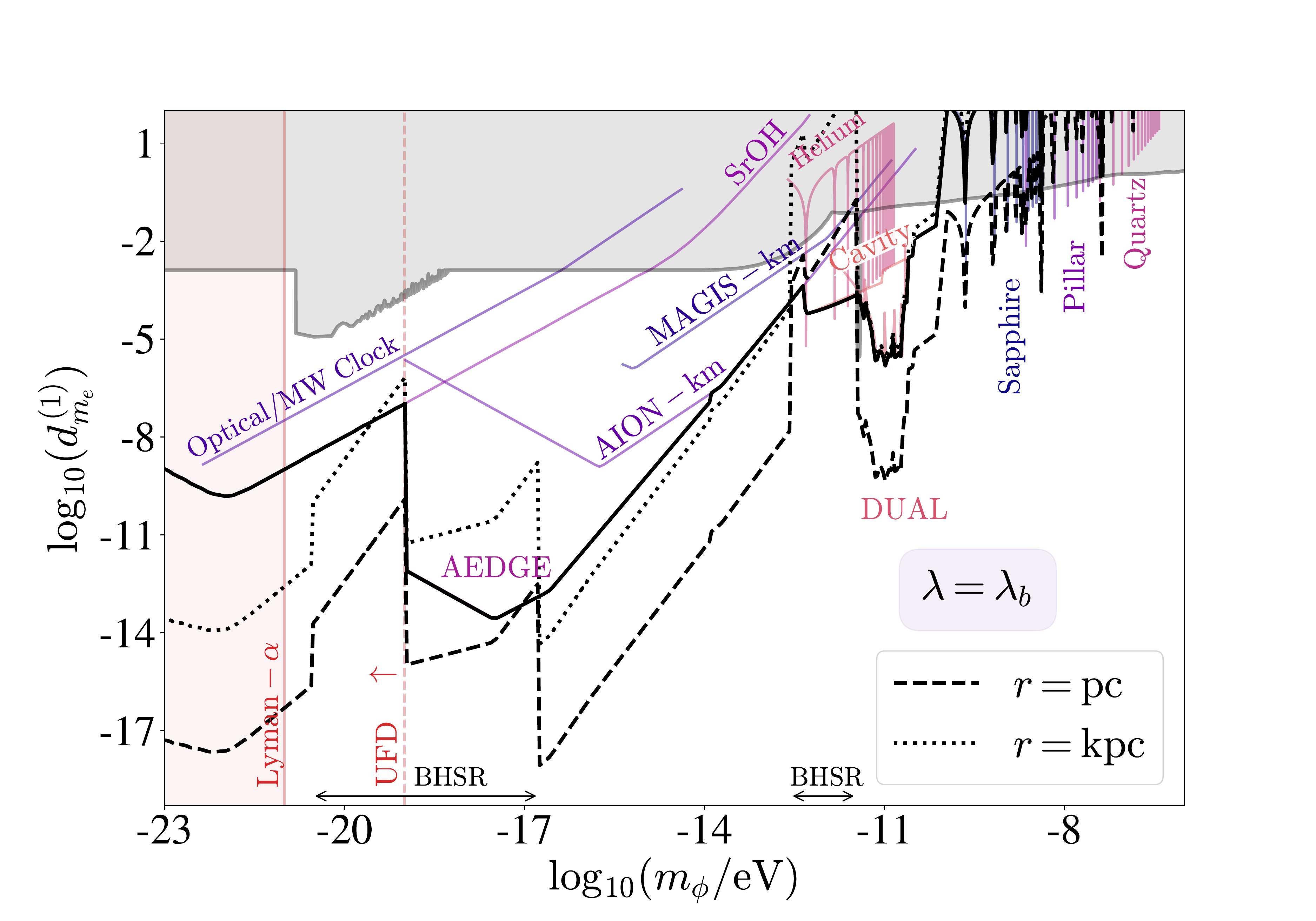}}
    \caption{Bounds on \(d_{m_e}^{(1)}\)  as a function of $m_{\phi}$ from current experiments (left) and projections for future experiments (right). In both plots, the solid black lines trace the best bound for each mass. The dashed and dotted black lines represent the potential reach for detecting bosenovae at distances of $r = $ pc and kpc, respectively. We choose $\lambda = \lambda_{b}$ in Eq.~\eqref{eq:lambdaB}, as explained in the text. Current experiments consist of various detector types. \textbf{Clocks:} Frequency comparison between $^{171}$Yb optical lattice clock and $^{133}$Cs fountain microwave clock (Yb/Cs) \cite{2022PhRvL.129x1301K}, \textbf{Optical cavities:} H-maser comparison with a Si cavity (H/Si) \cite{Kennedy:2020bac}, comparison between Cs clock and a cavity (Cs/Cav) \cite{Tretiak:2022ndx}, and comparison of a H-maser and sapphire oscillator with a quartz oscillator (H/Quartz/Sapphire) \cite{Campbell:2020fvq}, \textbf{Optical interferometers:} Unequal delay interferometer experiment (DAMNED) \cite{Savalle:2020vgz}, co-located Michelson interferometers (Holometer) \cite{Aiello:2021wlp}, \textbf{Spectroscopy:} Molecular iodine spectroscopy (I$_2$) \cite{Oswald:2021vtc}, comparison between $^{87}$Rb hyperfine transition and quartz mechanical oscillator (Rb/Quartz) \cite{Zhang:2022ewz}, \textbf{Equivalence principle tests:} Searches for EP violation (E$\ddot{\rm{o}}$t-Wash \cite{Hees:2018fpg} and MICROSCOPE \cite{Berge:2017ovy}), \textbf{Mechanical Oscillators:} Resonant mass detector (AURIGA) \cite{Branca:2016rez}, \textbf{Fifth Force tests:} Fifth force searches looking for deviations from the gravitational inverse square law \cite{Murata:2014nra}, and \textbf{Gravitational wave detectors:} (GEO 600 \cite{Vermeulen:2021epa}, LIGO O3 \cite{Fukusumi:2023kqd}). Future experiments include Optical/MW clock comparison \cite{Arvanitaki:2014faa}, SrOH molecular spectroscopy \cite{SrOH2021}, atom interferometery (terrestrial MAGIS \cite{2021QS&T....6d4003A}, and space-based AEDGE and AION \cite{Badurina:2021rgt}), resonant-mass detectors (DUAL \cite{Arvanitaki:2015iga}) and other mechanical resonators (Sapphire, Pillar, Quartz and Superfluid Helium \cite{Manley:2019vxy}). 
    Ultra-faint dwarf galaxies (UFDs)~\cite{Dalal:2022rmp} and Lyman-$\alpha$~\cite{Irsic:2017yje,Rogers:2020ltq} also constrain ULDM. 
    Finally, the regions of the parameter space corresponding to the black hole superradiance (BHSR) bounds~\cite{Baryakhtar:2020gao,Unal:2020jiy} are displayed at the bottom of the plots by the arrows (note the discontinuity in $\lambda_b$ in this range).} 
    \label{fig:dme1_current}
\end{figure*}

\begin{figure*}
    \centering
    
    \makebox[ \paperwidth]{\hspace{-3.5cm}\includegraphics[width = 0.6\linewidth]{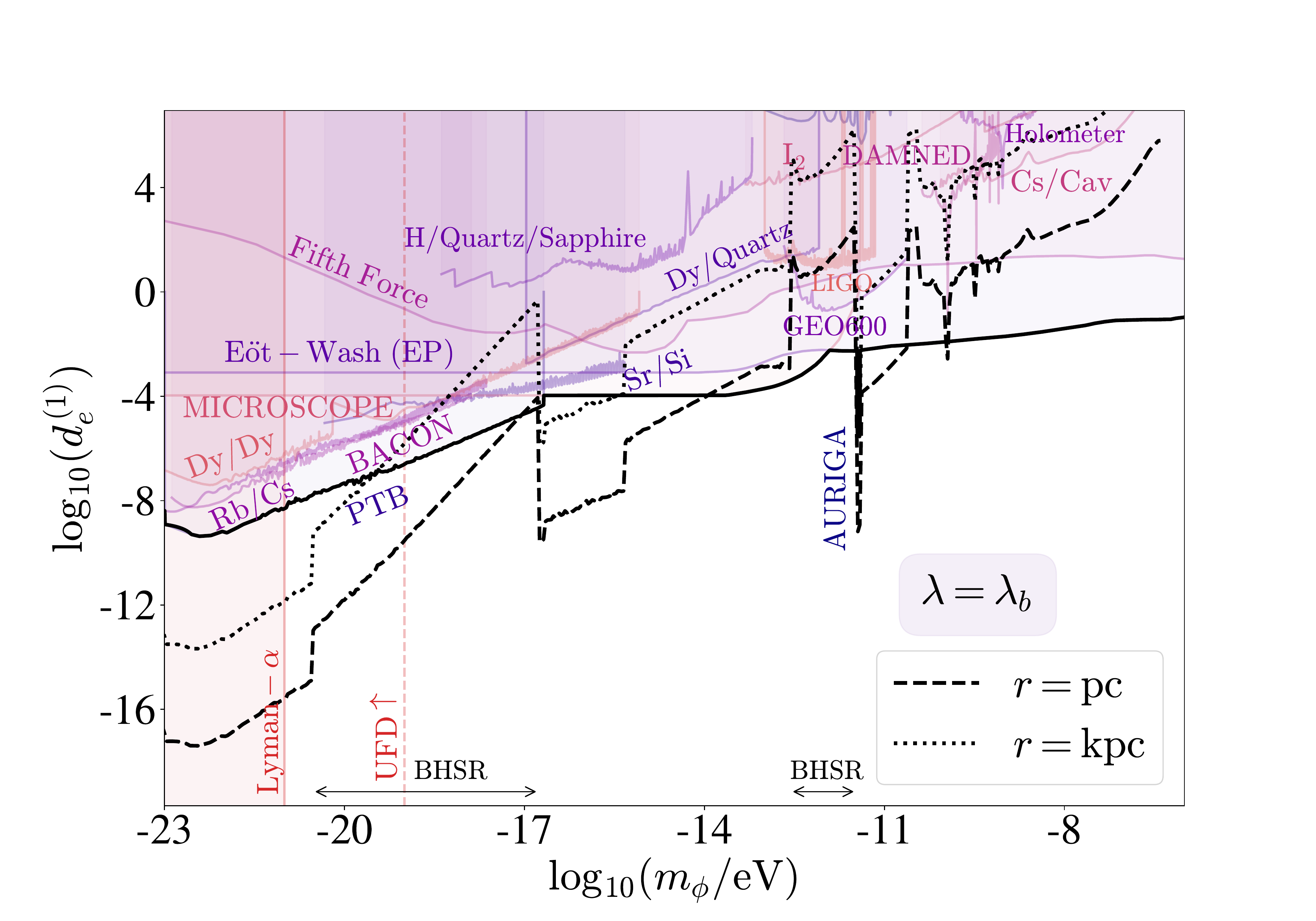}
    \hspace{-0.9cm}
    \includegraphics[width = 0.6\linewidth]{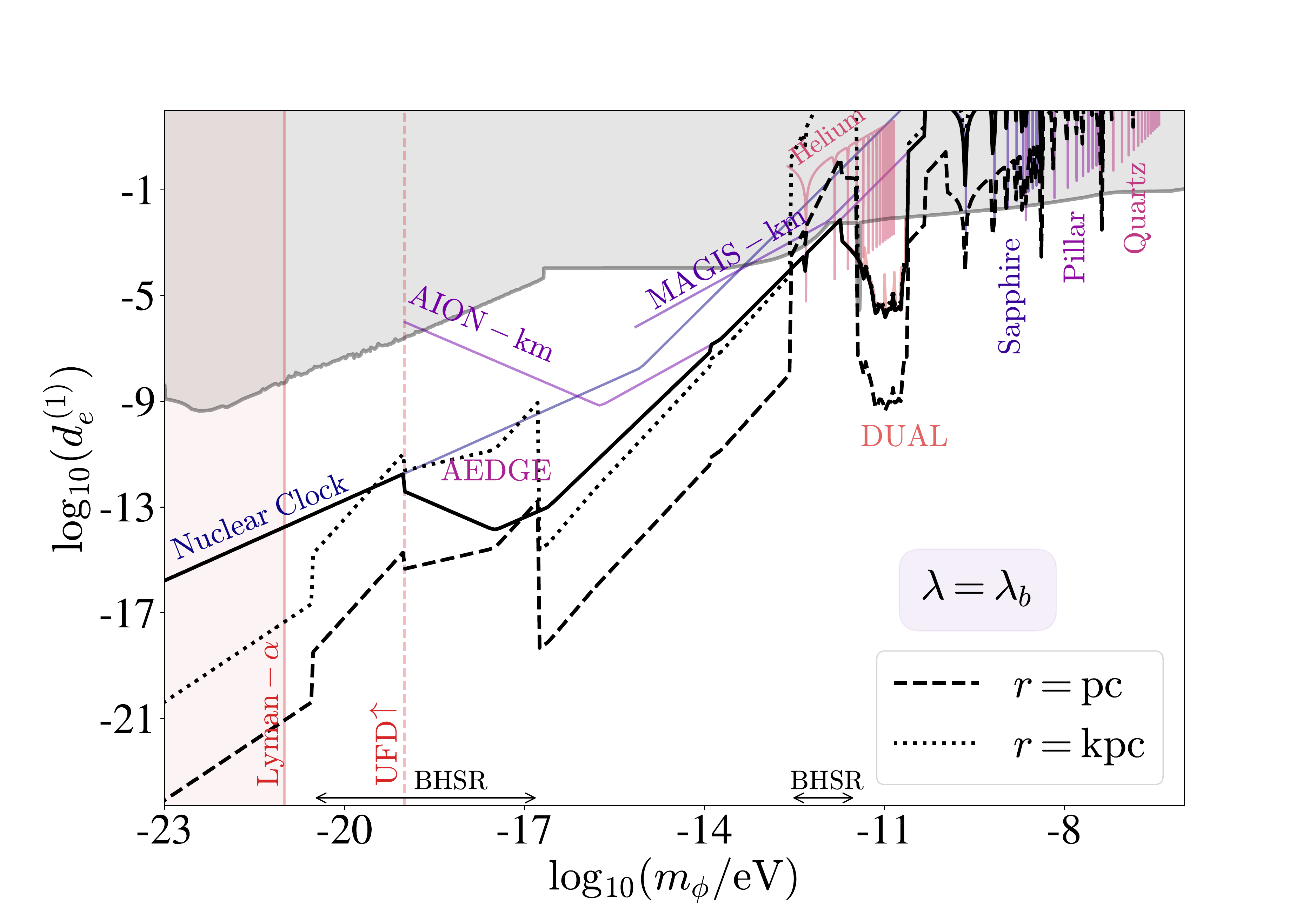}}
    \caption{Bounds on \(d_{e}\) as a function of $m_{\phi}$ from current experiments (left) and projections for proposed experiments (right). In both plots, the solid black lines trace the best bound for each mass. The dashed and dotted black lines represent the potential reach for detecting bosenovae at distances of $r = $ pc and kpc, respectively. We choose $\lambda = \lambda_{b}$ in Eq.~\eqref{eq:lambdaB}, as explained in the text.  Current experiments include the following. \textbf{Clocks:} Frequency ratios of $^{27}$Al$^+$, $^{171}$Yb and $^{87}$Sr optical clocks (BACON) \cite{Beloy:2020tgz}, frequency ratios of single-ion $^{171}$Yb$^+$ clock and $^{87}$Sr optical lattice clock (PTB) \cite{Filzinger:2023zrs}, ratio of frequencies of atomic clocks based on $^{171}$Yb$^+$ and $^{87}$Sr (Yb+/Sr) \cite{Sherrill:2023zah}, and dual $^{133}$Cs/$^{87}$Rb atomic fountain clock frequency ratio (Rb/Cs) \cite{Hees:2016gop}; \textbf{Optical cavities} Frequency comparisons between Strontium optical clock and Silicon cavity (Sr/Si) \cite{Kennedy:2020bac}, atomic spectroscopy in Cesium vapor with Fabry-Perot cavity locked to laser (Cs/Cav) \cite{Tretiak:2022ndx} and frequency comparision of Hydrogen maser and sapphire oscillator with quartz oscillator (H/Quartz/Sapphire) \cite{Campbell:2020fvq}; \textbf{Optical interferometers} Three-arm Mach-Zender interferometer (DAMNED) \cite{Savalle:2020vgz} and co-located Michelson interferometers (Holometer) \cite{Aiello:2021wlp}; \textbf{Spectroscopy} Spectroscopic experiments of molecular iodine (I$_2$) \cite{Oswald:2021vtc}, frequency comparison of $^{164}$Dy with quartz oscillator (Dy/Quartz) \cite{Zhang:2022ewz}, precision spectroscopy measurements involving two isotopes of dysprosium (Dy/Dy) \cite{VanTilburg:2015oza}; \textbf{Equivalence principle tests} Tests of equivalence principle violation by (E$\ddot{\rm{o}}$t-Wash) \cite{Hees:2018fpg} and (MICROSCOPE) \cite{Berge:2017ovy}; \textbf{Mechanical resonators} Resonant mass detector (AURIGA) \cite{Branca:2016rez}; and \textbf{Gravitational wave detectors} (GEO 600) \cite{Vermeulen:2021epa} and (LIGO O3) \cite{Fukusumi:2023kqd}. Future experiments include Thorium Nuclear Clock projections \cite{Antypas:2022asj}, space-based atom interferometers (AEDGE and AION) \cite{Badurina:2021rgt}, terrestrial atom interferometers  (MAGIS) \cite{2021QS&T....6d4003A}, resonant-mass detectors DUAL \cite{Arvanitaki:2015iga}, and mechanical resonators (Sapphire, Pillar, Quartz and Superfluid Helium)~\cite{Manley:2019vxy}.
    Ultra-faint dwarf galaxies (UFDs)~\cite{Dalal:2022rmp} and Lyman-$\alpha$~\cite{Irsic:2017yje,Rogers:2020ltq} also constrain ULDM. 
    Finally, the regions of the parameter space corresponding to the black hole superradiance (BHSR) bounds~\cite{Baryakhtar:2020gao,Unal:2020jiy} are displayed at the bottom of the plots by the arrows (note the discontinuity in $\lambda_b$ in this range).}
    \label{fig:de1_current}
\end{figure*}

\begin{figure*}
    \centering
    
    \makebox[ \paperwidth]{\hspace{-3.5cm}\includegraphics[width = 0.6\linewidth]{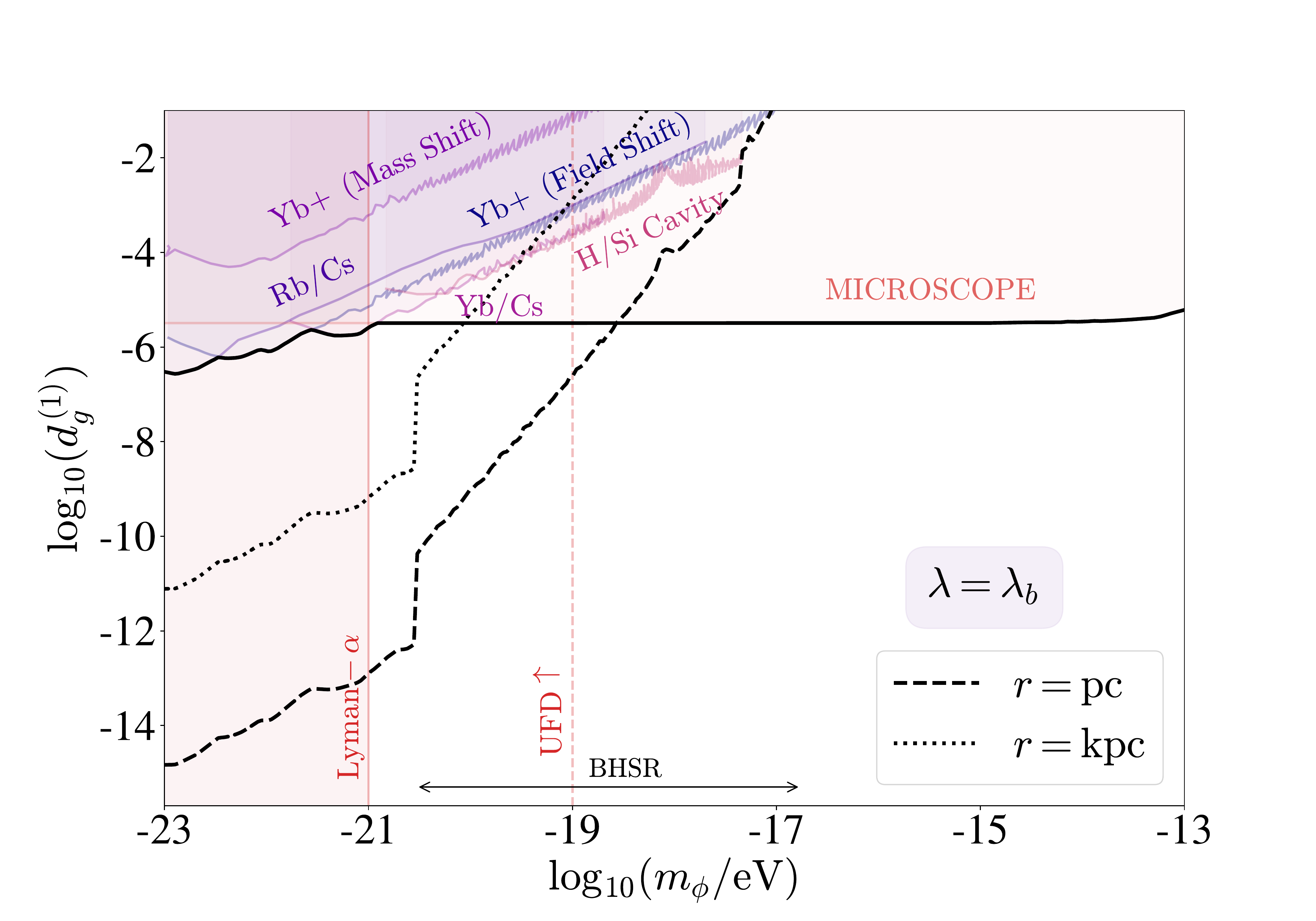}
    \hspace{-0.9cm}
    \includegraphics[width = 0.6\linewidth]{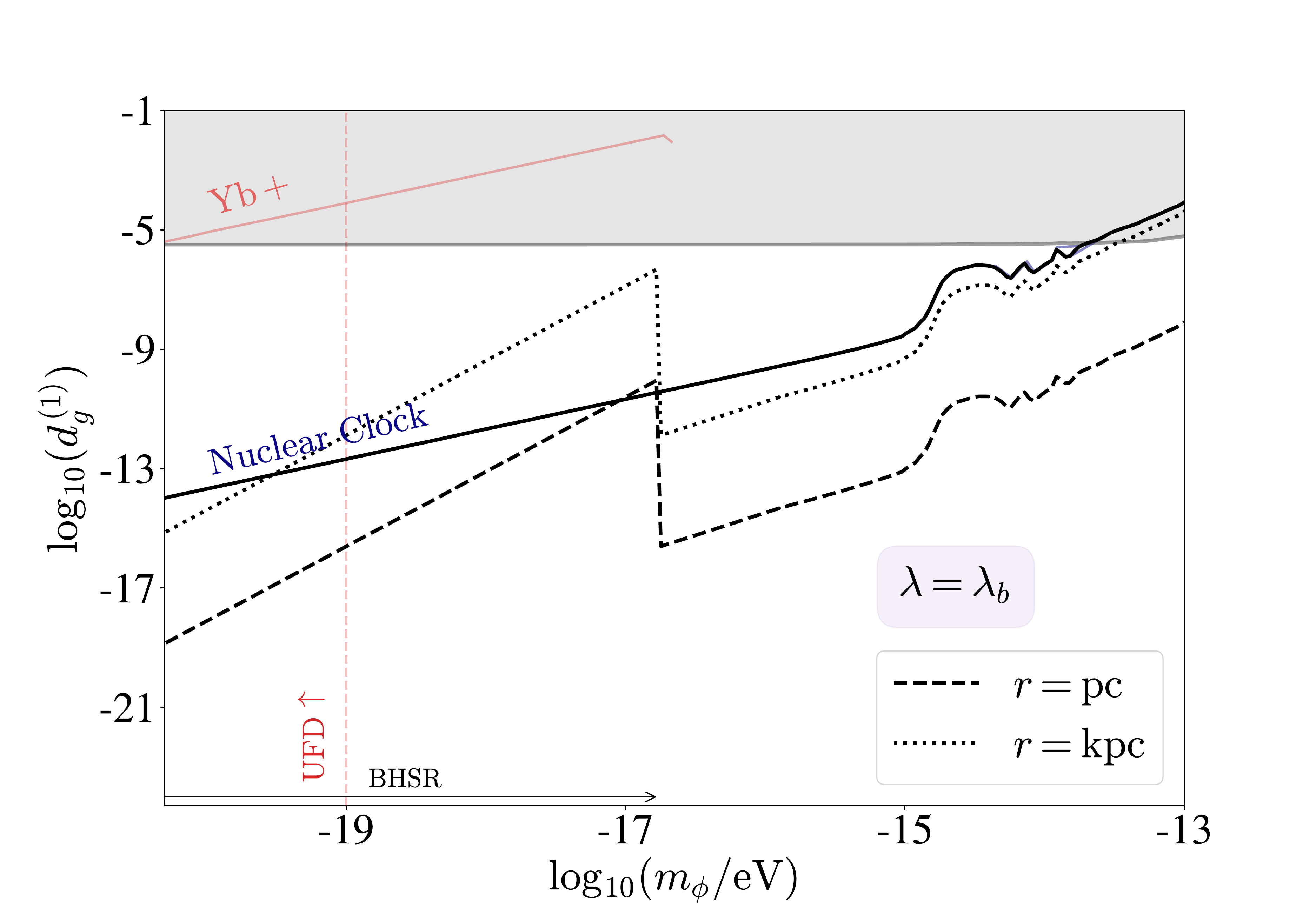}}
    \caption{Bounds on \(d_{g}\) as a function of $m_{\phi}$ from current experiments (left) and projections for proposed experiments (right). In both plots, the solid black lines traces the best bound for each mass. The dashed and dotted black lines represent the potential reach for detecting bosenovae at distances of $r = $ pc and kpc, respectively. We choose $\lambda = \lambda_{b}$ in Eq.~\eqref{eq:lambdaB}, as explained in the text. Experiments that are sensitive to $d_g^{(1)}$ include \textbf{Clocks:} $^{171}$Yb lattice clock - $^{133}$Cs microwave clock comparison (Yb/Cs) \cite{2022PhRvL.129x1301K} and frequency comparison in dual rubidium-caesium cold atom clock (Rb/Cs) \cite{Hees:2016gop}, \textbf{Optical cavities:} H-maser comparison with a Si cavity (H/Si) \cite{Kennedy:2020bac}, and \textbf{Equivalence principle tests:} MICROSCOPE \cite{Berge:2017ovy}, while future experiment projections are for the $^{229}$Th Nuclear Clock \cite{Antypas:2022asj} and $^{171}$Yb$^+$ ion clock~\cite{Banerjee:2023bjc}.
    Ultra-faint dwarf galaxies (UFDs)~\cite{Dalal:2022rmp} and Lyman-$\alpha$~\cite{Irsic:2017yje,Rogers:2020ltq} also constrain ULDM. 
    Finally, the regions of the parameter space corresponding to the black hole superradiance (BHSR) bounds~\cite{Baryakhtar:2020gao,Unal:2020jiy} are displayed at the bottom of the plots by the arrows (note the discontinuity in $\lambda_b$ in this range).}
    \label{fig:dg1_current}
\end{figure*}

Figs.~\ref{fig:dme1_current}, \ref{fig:de1_current}, and \ref{fig:dg1_current} show the reach of the \(d^{(1)}_{m_e}\),  \(d^{(1)}_{e}\), and \(d^{(1)}_{g}\), respectively, as a function of \(m_{\phi}\). We display the reach for both current experiments (left) and proposed future experiments (right). For this parameter space, there are two additional parameters: the self coupling \(\lambda\) and bosenova distance \(r\). As before, each figure shows two benchmark distances of \(r = 1\) pc and \(r = 1\) kpc.

Since \(\rho_{*} \propto 1/\lambda\,\) (see Eq.~\eqref{eq:density_wavespread}), smaller values of \(\lambda\) will increase detection prospects. Taking into account all of the constraints on $\lambda$ laid out in Section~\ref{ssec:lambdaconstraints}, we choose 
values of $\lambda$ at each mass which will maximize the size of the signal. The benchmark $\lambda_b$ is chosen therefore to trace right above the lower bounds on $\lambda$, as shown by the black dashed line in Fig.~\ref{fig:lambda_constraints}. This benchmark choice is defined piece-wise by
\begin{align} \label{eq:lambdaB}
    \lambda_{\text{b}} = 
    \begin{cases}
       \displaystyle 3\cdot 10^{-73} \bigg(\frac{m_{\phi}}{10^{-13}\,{\rm eV}}\bigg)^6 
       & \displaystyle 10^{-13} \lesssim \frac{m_{\phi}}{\rm eV} \lesssim 10^{-12} \\
       \\
       \displaystyle 10^{-90} \bigg(\frac{m_{\phi}}{5\cdot 10^{-21}\,{\rm eV}}\bigg)^4 
       & \displaystyle 10^{-21} \lesssim \frac{m_{\phi}}{\rm eV} \lesssim 10^{-17} \\
       \\
       \displaystyle 10\,\lambda_{\text{BH}} & {\rm elsewhere}
    \end{cases}\,.
\end{align}
For the regions in Fig.~\ref{fig:lambda_constraints} not dominated by black hole superradiance bounds, we take \(\lambda = 10\,\lambda_{\text{BH}}\) to safely ignore relativistic effects that would arise and modify the dynamics of the bosenovae. We emphasize that this benchmark is chosen by hand based on existing constraints, and the precise form was not derived from any theoretical considerations. 

\begin{figure*}
    \centering
    \makebox[ \paperwidth]{\hspace{-3.5cm}\includegraphics[width = .6\linewidth]{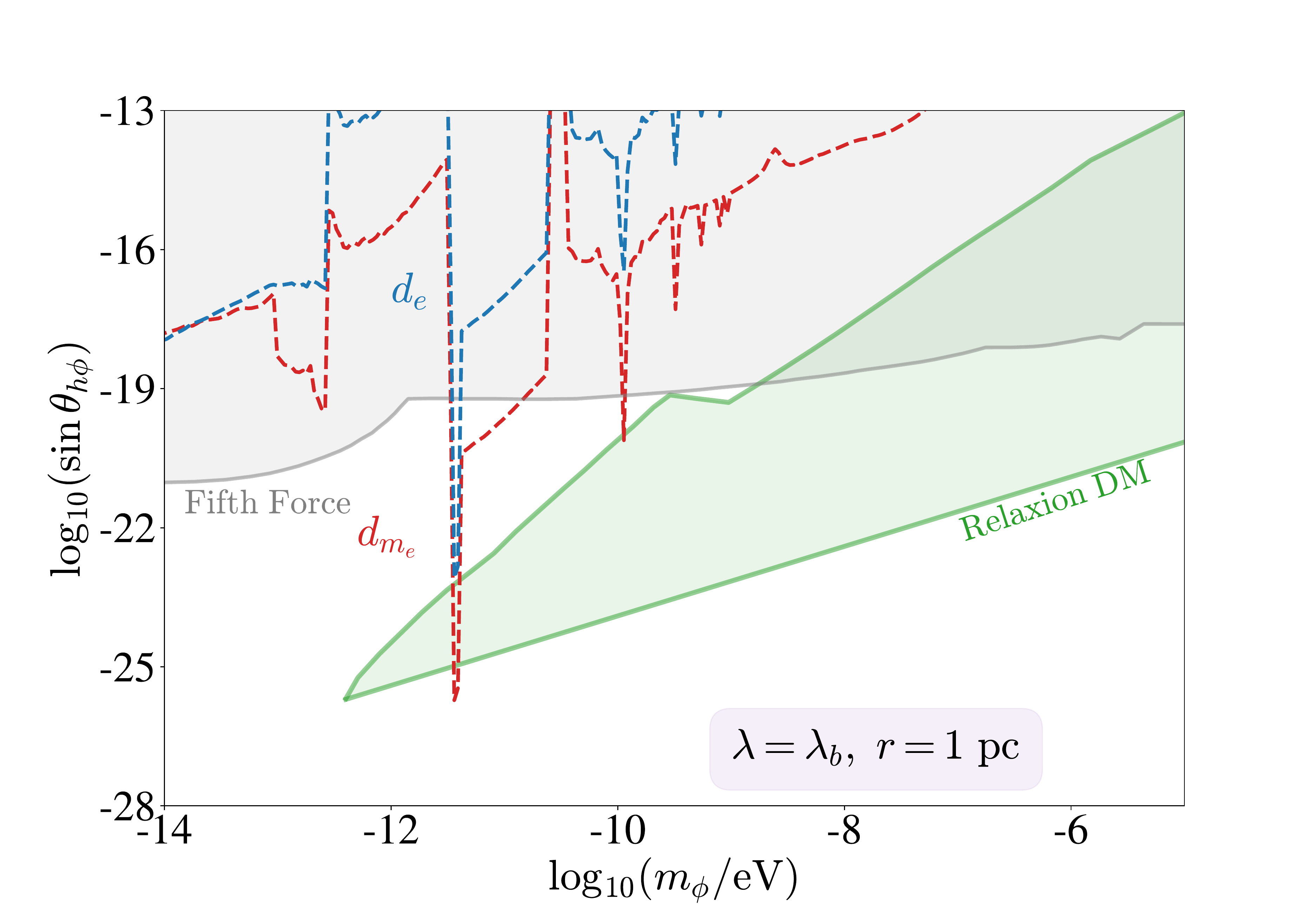}
    \hspace{-0.9cm}
    \includegraphics[width = .6\linewidth]{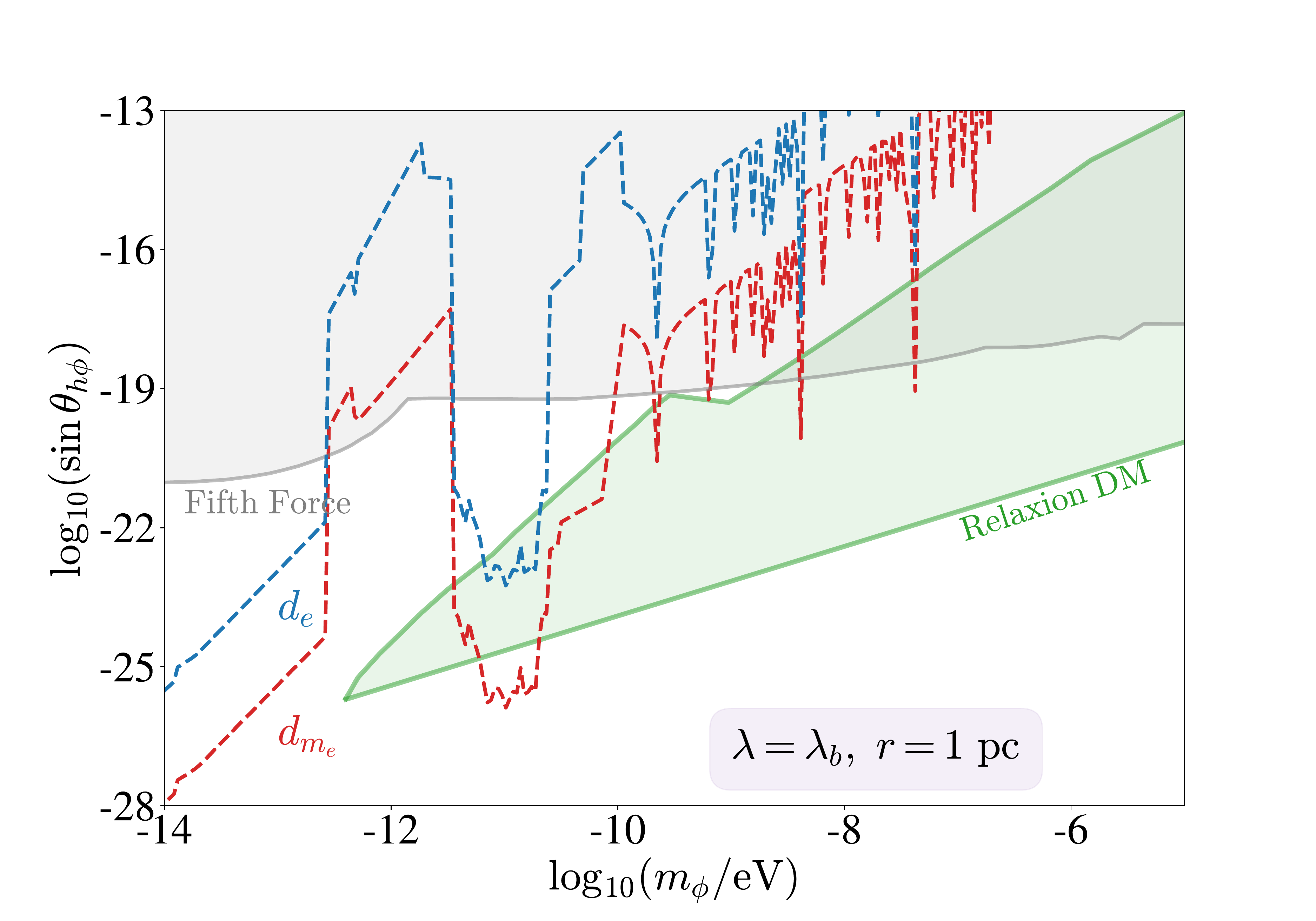}}
    \caption{The reach for bosenova search compared to the relaxion DM parameter space (green shaded region~\cite{Chatrchyan:2022dpy}), for current experiments (left) and future experiments (right). 
    The strongest current constraints, derived from EP and fifth force experiments, are represented by the light gray region. The potential reach for bosenova searches for $d_{m_e}$ and $d_e$ are shown by the red and blue dashed lines and derived using Eq.~\eqref{eq:relaxionmixing} from the experiments listed in Figs.~\ref{fig:dme1_current} and \ref{fig:de1_current} (respectively). The bosenovae were taken to be at a benchmark distance of $r = $ pc.}
    \label{fig:relaxion_plot}
\end{figure*}

Finally, we illustrate a dedicated relaxion parameter space 
in Fig.~\ref{fig:relaxion_plot}. We display the parameter space in terms of the Higgs-relaxion mixing parameter, $\sin{\theta_{h \phi}}$, which is related to the dilatonic couplings via Eq.~(\ref{eq:relaxionmixing}). With upcoming experiments probing $d_e$ and $d_{m_e}$, there will be significant potential to probe well-motivated relaxion DM parameter space. 

Finally, we note that although the parameter space $m_\phi\lesssim 10^{-19}\,{\rm eV}$ is already constrained by astrophysics and cosmology (especially UFDs and Lyman-$\alpha$), laboratory searches are still useful and complementary in this range. Laboratory and astrophysical probes are based on very different assumptions and therefore provide useful confirmation of one another. Furthermore, in models where the ULDM scalar field is a subleading fraction $f\lesssim 0.1$ of the total DM density, astrophysical constraints can disappear completely, whereas laboratory searches typically reduce sensitivity by only $\sqrt{f}$. In fact, in the present context, the total ULDM density only affects the rate of bosenovae, but not the size of a given bosenova signal if it does occur.

\section{Conclusions and Outlook}
\label{sec:conclusion}

ULDM can form compact objects that eventually collapse and explode, resulting in transient burst emissions of relativistic scalar fields. This can lead to distinct observational signatures compared to Galactic DM. We have demonstrated that current experiments searching for DM couplings to photons, electrons, and gluons may already be sensitive to such bosenova signals. Upcoming experiments and technology, including nuclear clocks as well as space-based interferometers, will be able to sensitively probe orders of magnitude in the ULDM coupling-mass parameter space.

The analysis put forth in our work is general and the methodology can be readily applied to other astrophysical sources of relativistic scalar fields. 
Scalar particle emission can generally originate from hot and dense environments\footnote{For example, such as that found in astrophysical settings of neutron star mergers~\cite{Harris:2020qim, Fiorillo:2021gsw}.}.
Our work opens up new avenues for multimessenger astronomy associated with new physics as well.

Since the properties of the bosenova are linked to $m_{\phi}$ and $\lambda$, other transient sources could have a distinct dependence on these parameters, and therefore lead to different detection prospects. In particular, in the event of a bosenova detection, the pattern of peaks in the emission spectrum may encode important information about the underlying UV physics, which is inaccessible in an ordinary DM search. 
A dedicated study of the emission spectrum of bosenovae in theories with different scalar field potentials is therefore highly motivated. 

\appendix 

\section*{Acknowledgements}

We thank Abhishek Banerjee and John Ellis for useful discussions.
This work was supported in part by the NSF QLCI Award OMA - 2016244, NSF Grant PHY-2012068, and the European Research Council (ERC) under the European Union’s Horizon 2020 research and innovation program (Grant Number 856415). The work of JE and VT was supported by the World Premier International Research Center Initiative (WPI), MEXT, Japan, and by the JSPS KAKENHI Grant Numbers 21H05451 (JE), 21K20366 (JE) and 23K13109 (VT).
This research was supported in part by the INT's U.S. Department of Energy grant No. DE-FG02- 00ER41132.

\bibliographystyle{JHEP}
\bibliography{ref}

\end{document}